 \RequirePackage{fix-cm}
\documentclass[smallcondensed]{svjour3}     % onecolumn (ditto)
\usepackage{mathrsfs}
  \usepackage{amsmath}
  \usepackage{amssymb}
 \usepackage{bm}
 \usepackage{arydshln}
  \usepackage{subfig}
   \usepackage{natbib}
   \usepackage{graphicx}
   \usepackage{float}
\usepackage{epstopdf, epsfig}
\usepackage{physics}
\usepackage{color}
\usepackage{stmaryrd}
\usepackage{natbib}
\usepackage{graphics}
\usepackage{appendix}
 \usepackage{algorithmic}
\usepackage[ruled,vlined]{algorithm2e}

\title{A stochastic SPOD-Galerkin model for broadband turbulent flows%\thanks{Grants or other notes
}
\author{Tianyi Chu \and Oliver T. Schmidt}

\institute{T. Chu \at
            Department of Mechanical and Aerospace
 Engineering, Jacobs School of Engineering, UCSD, 9500 Gilman Drive, La
 Jolla, CA 92093-0411, USA \\
              \email{tic173@eng.ucsd.edu}           %  \\
          \and
         O. Schmidt \at
            Department of Mechanical and Aerospace
 Engineering, Jacobs School of Engineering, UCSD, 9500 Gilman Drive, La
 Jolla, CA 92093-0411, USA \\
              \email{oschmidt@ucsd.edu} 
}

\date{Received: date / Accepted: date}

\begin{document}

\maketitle
 
 \begin{abstract}

The use of spectral proper orthogonal decomposition (SPOD) to construct low-order models for broadband turbulent flows is explored. The choice of SPOD modes as basis vectors is motivated by their optimality and space-time coherence properties for statistically stationary flows. This work follows the modeling paradigm that complex nonlinear fluid dynamics can be approximated as stochastically forced linear systems. The proposed stochastic two-level SPOD-Galerkin model governs a compound state consisting of the modal expansion coefficients and forcing coefficients. In the first level, the modal expansion coefficients are advanced by the forced linearized Navier-Stokes operator under the linear time-invariant assumption. The second level governs the forcing coefficients, which compensate for the offset between the linear approximation and the true state. At this level, least squares regression is used to achieve closure by modeling nonlinear interactions between modes. 
The statistics of the remaining residue are used to construct a dewhitening filter that facilitates the use of white noise to drive the model. If the data residue is used as the sole input, the model accurately recovers the original flow trajectory for all times. If the residue is modeled as stochastic input, then the model generates surrogate data that accurately reproduces the second-order statistics and dynamics of the original data. The stochastic model uncertainty, predictability, and stability are quantified analytically and through Monte Carlo simulations. The model is demonstrated on large eddy simulation data of a turbulent jet at Mach number $M=0.9$ and Reynolds number $\mathrm{Re}_D\approx 10^6$.
\end{abstract}

% \PACS{PACS code1 \and PACS code2 \and more}
% \subclass{MSC code1 \and MSC code2 \and more}

\section{Introduction}
High-fidelity numerical simulations of turbulent flows have become common practice in engineering and science. Their computational cost, however, is often prohibitive for optimization, and their time-to-solution hinders real-time control. Reduced-order models (ROMs) are routinely employed to reduce computational complexity in this context.

% \Red{concept 1 : POD-Galerkin}

Among the most successful models in fluid dynamics are POD-Galerkin models that seek a low-order representation of the flow dynamics in the space spanned by a modal basis obtained from proper orthogonal decomposition (POD). Most commonly, the so-called method of snapshots introduced by \citet{sirovich1987turbulence} is used to obtain a basis of space-only POD modes and temporal expansion coefficients. This most popular flavor of POD is computed from the eigendecomposition of the spatial cross-correlation tensor. The resulting POD modes optimally represent the data in terms of a spatial inner product. POD-based Galerkin models leverage the orthogonality of POD in the spatial inner product to obtain low-order models that govern the temporal evolution of the POD expansion coefficients. \citet{holmes1996turbulence} and \citet{rowley2017model} summarized and illustrate this approach. POD-Galerkin models have been successfully applied to model the wall region of incompressible turbulent boundary layers by \citet{aubry1988dynamics} and later, with a focus on low wave number phenomena of turbulence generation, by \citet{holmes1997low}. \citet{noack2003hierarchy} proposed a hierarchy of POD-Galerkin models for viscous cylinder wakes and introduced a shift mode that accounts for the mean field correction. For compressible flows governed by the compressible governing equations, POD-Galerkin modeling is significantly complicated by the additional energy equation and the occurrence of triple products of the state variables, which may be defined in either primitive or conservative form. It was observed by \citet{rempfer2000low} that POD-Galerkin models of general compressible flows exhibit nonphysical instability. There are several previous attempts to overcome these difficulties for moderate Reynolds number flows. In pioneering works by \citet{rowley2001dynamical,rowley2002modeling}, the authors implemented POD-Galerkin models for compressible cavity flows for control purposes.
Later, \citet{rowley2004model} applied simplified Navier-Stokes equations to obtain a quadratic POD-Galerkin ROM that is valid for isentropic, cold flows at moderate
Mach numbers. For linearized compressible flows, \citet{barone2009stable} devised an inner symmetry product that yields stable models. Expanding on this work, \citet{kalashnikova2011stable,kalashnikova2012efficient} proposed stability-preserving model reduction techniques for non-linear systems. All of the work discussed so far is based on standard space-only POD computed from the method of snapshots.

As an alternative to POD-Galerkin techniques, which require knowledge of the governing equations, purely data-driven models are commonly used in the atmospheric sciences. Inverse stochastic models, for example, approximate the temporal evolution of the POD expansion coefficients directly from the input data. The simplest type of inverse stochastic models is the so-called linear inverse model (LIM) introduced by
\citet{penland1989random,penland1996stochastic}.
The underlying assumption of this model is that
a nonlinear dynamical system can be modeled as a deterministic linear system that is driven by stochastic forcing. In practice, the linear operator which governs the evolution is obtained by linear regression, and the regression residue is estimated as white noise.
Within the LIM framework, additive noise is used to model both the nonlinear flow physics and process noise. LIMs have successfully been applied in climate and weather prediction, for example of the El Ni\~{n}o-Southern Oscillation \citep{penland1995optimal,johnson2000empirically}, and other sea surface temperature anomalies \citep{penland1998prediction}. 
Later, as a generalization of LIMs, \citet{kondrashov2005hierarchy,kravtsov2005multilevel} introduced the multiple polynomial regression (MPR), which expands on the purely linear system assumption of LIMs by assuming a polynomial form for the expansion coefficients to better approximate nonlinear dynamics. Expanding on this idea, the authors further proposed the use of a multi-level regression (MLR) model, which inflates the MPR model by a hierarchy of additional levels until the residue can be modeled as white noise. The additional levels are used to account for the nonlinearity of the system and serial correlations in the residue. More recently, \citet{kondrashov2015data} generalized the MLR models by introducing additional implicit variables. All of the above inverse stochastic models are applied to approximate the nonlinear dynamics of the expansion coefficients of space-only POD modes. More recently, \citet{chekroun2017data} proposed a model for the time-domain expansion coefficients of data-adaptive harmonic modes, which to the best understanding of the present authors correspond to spectral POD (SPOD) modes, by specifying the main-level polynomial form in MLRs as a Stuart-Landau oscillator.

% As an application, the multi-level inverse models can be applied to approximate the nonlinear forcing of the linearized compressible Navier-Stokes equations.

% \Red{concept 3 : SPOD}
% Here, we propose a different modeling approach based on spectral POD (SPOD). 

SPOD, the frequency-domain variant of POD, was first conceptually introduced by \citet{lumley1967structure,lumley2007stochastic}, and is computed as the eigendecomposition of the cross-spectral density tensor under the assumption of ergodicity, that is, for statistically stationarity flows.
SPOD decomposes the data set into energy-ranked, monochromatic modes \citep{towne2018spectral,schmidt2018spectral}. The resulting modes evolve coherently in both space and time, and optimally represent the second-order space-time flow statistics. SPOD has been applied to a variety of turbulent flows, including wall-bounded shear flows such as boundary layers \citep{tutkun2017lumley}, as well as bluff-body flows such as the wake behind a disk \citet{tutkun2008three}. For turbulent jets, the use of SPOD was pioneered by \citet{glauser1987coherent}. Later, different authors linked coherent structures identified by SPOD to concepts from linear stability theory for both
experimental data \citep{gordeyev2000coherent,gudmundsson2011instability,cavalieri2013wavepackets} and large eddy simulation (LES) data \citep{schmidt2017wavepacket,schmidt2017wavepackets,towne2017acoustic,towne2018spectral}. 

In this work, we propose an SPOD-Galerkin two-level model for stochastic modeling of fully turbulent flows under the linear-time invariant (LTI) approximation. The use of SPOD modes as the basis for the model is motivated by: (i) their optimality for statistically stationary data, and (ii) by the prospect that SPOD modes can be modeled as eigenvectors of the resolvent operator \citep{towne2018spectral}. The latter property has two important implications. First, the choice of modal basis is consistent with the LTI approximation, that is, that of forced linear dynamics, and second, the potential to enable future models that use resolvent modes as basis (note that classical POD modes are data-driven and cannot readily be predicted by theory).

The goal of this work is to devise a ROM that can accurately reproduce a known flow trajectory, or, when forced stochastically, generates surrogate data which reproduces the second-order statistics and dynamics of chaotic, high-Reynolds number flows. Applications of the latter capability include Monte Carlo-based prediction under uncertainty.The key idea of this framework is to approximate the nonlinear fluid dynamics obtained from SPOD-Galerkin projection by a stochastically forced LTI system. The resulting forced ROM is considered as the first level of the model, which advances the state coefficients in time by the linearized Navier-Stokes operator. Due to the convective nonlinearity of the Navier-Stokes equations, white-in-time noise is generally not adequate as the sole stochastic input to linearized flow models. \citet{zare_jovanovic_georgiou_2017}, for example, presented a modeling framework for turbulent channel flow which obtains colored-in-time stochastic excitation to account for second-order flow statistics by solving a convex optimization problem.
Here, we take a different approach by incorporating the forcing coefficients as part of the state to offset the difference between the linear approximation and the true, nonlinear system dynamics.
Finally, linear dependence between the evolution of the state and forcing coefficients is established by means of least squares regression. Closure of the model is achieved, and can be tested \emph{a priori}, if the second level residue takes on the characteristics of white noise. 

% \Red{
% Compared to classical Galerkin-POD models, which are usually demonstrated on low Reynolds-number flows or use eddy viscosity models and therefore do not model the finer scales and stochastic nature of turbulence, this proposed 2-level model is applicable for more complex flows.
%  }

Following brief introductions of Galerkin projection and SPOD in \S \ref{background}, the stochastic two-level SPOD-Galerkin ROM is derived in \S \ref{SPOD-Galerkin ROM}, including residue modeling and uncertainty quantification. The performance of the SPOD-Galerkin two-level model is demonstrated in \S \ref{example} using the LES data of the axisymmetric component of a turbulent jet at Mach number $M=0.9$ and jet-diameter Reynolds number $\mathrm{Re}_D\approx 10^6$. By verifying that the second level residues are well approximated by white noise, model closure is confirmed. We show that the surrogate data obtained from our model accurately reproduces the second-order statistics and dynamics of the input data. The model uncertainty and stability are quantified by analytical means and Monte Carlo simulation. Finally, \S \ref{conclusion} summarizes and concludes the paper. 
%%%%%%%%%%%%%%%%%%%%%%%%%%%%%%%%%

% To offset the difference between the linear approximation and the true, nonlinear system, we further incorporate the forcing coefficients as part of the state in the second level. 

\section{Background} \label{background}

\subsection{Galerkin projection-based model order reduction for linearized Navier-Stokes equations} \label{galerkin}

We start by taking the Reynolds decomposition of the flow state $\vb*{q}\in \mathbb{C}^n$ into the temporal mean and fluctuating components $\overline{(\cdot)}$ and  $(\cdot)'$, respectively, as
\begin{align}
    \vb*{q}(t)=\bar{\vb*{q}}+\vb*{q}'(t). \label{flow_d}
\end{align}
As an integral measure of energy we define the inner product and associated norm
\begin{align}
    \left<\vb*{q}_1,\vb*{q}_2\right>_E&=\vb*{q}_1^*\vb*{W}\vb*{q}_2, \quad\text{and}\label{i_prod} \\
    \|\vb*{q}\|^2_E &= \left<\vb*{q},\vb*{q}\right>_E,  \label{norm}
\end{align}
respectively, where $\vb*{W}$ is a diagonal positive-definite weight matrix, and $(\cdot)^*$ denotes the Hermitian transpose.

% Suppose the dynamics of $\vb*{q}'$ are governed, in general, by
% \begin{align}
% \dv{}{t} \vb*{q}'=\vb*{g}(\vb*{q}',t;\bar{\vb*{q}}).\label{g}     
% \end{align}
% The reduced-order dynamics
% \begin{align}
%   \vb*{P} \dv{}{t} \vb*{q}'=\dv{}{t} \widetilde{\vb*{q}}= \vb*{V} \dv{}{t} \vb*{a}=\vb*{P}\vb*{g}(\widetilde{\vb*{q}},t;\bar{\vb*{q}}) \label{Pq}   
% \end{align}
% are obtained by projecting equation (\ref{g}) onto the subspace $\mathcal{S}$ and substituting $\widetilde{\vb*{q}}$ for $\vb*{q'}$. Left-multiplying (\ref{Pq}) by $\vb*{V}^* \vb*{W}$ yields
% \begin{align}
%     \vb*{V}^* \vb*{W} \vb*{V} \dv{}{t} \vb*{a}= \vb*{V}^* \vb*{W}\vb*{g}(\vb*{V}\vb*{a},t;\bar{\vb*{q}})
% \end{align}
% and the final form of the (weighted) Galerkin ROM that governs the evolution of the temporal expansion coefficients is obtained as
% \begin{align}
%     \ \dv{}{t} \vb*{a}=\left(\vb*{V}^* \vb*{W} \vb*{V}\right)^{-1} \vb*{V}^* \vb*{W}\vb*{g}(\vb*{V}\vb*{a},t;\bar{\vb*{q}}). \label{red_order}  
% \end{align}
% Equation (\ref{red_order}) is an $m$-dimensional first-order differential equation with initial condition $\vb*{a}(0)=\left(\vb*{V}^* \vb*{W} \vb*{V}\right)^{-1}\vb*{V}^*\vb*{W}\vb*{q}'(0)$.
% The classical form of the Galerkin ROM is obtained by choosing $\vb*{W}$ as the identity. 

% The most popular data decomposition techniques to in the context of Galerkin ROM are POD and dynamic mode decomposition (DMD, \citet{schmid2010dynamic}). Refer to the reviews by \citet{rowley2017model} and \citet{klus2018data} for more details.

As motivated above, we choose as the basis of the Galerkin ROM the first $M$ SPOD modes (at all $N_f$ frequencies),
\begin{equation}
  \vb*{V}= \left[
  \smash[b]{\underbrace{\begin{matrix}
  \vert &  &\vert \\
  \bm{\mathit{\psi}}_{\omega_1}^{(1)} & \cdots &\bm{\mathit{\psi}}_{\omega_1}^{(M)}\\
  \vert & &\vert
  \end{matrix}}_{\omega_1 }} \quad 
   \smash[b]{\underbrace{\begin{matrix}
  \vert & &\vert \\
  \bm{\mathit{\psi}}_{\omega_2}^{(1)} & \cdots &\bm{\mathit{\psi}}_{\omega_2}^{(M)}\\
  \vert & &\vert
  \end{matrix}}_{\omega_2 }} \quad  \cdots \quad 
   \smash[b]{\underbrace{\begin{matrix}
  \vert & &\vert \\
  \bm{\mathit{\psi}}_{\omega_{N_f}}^{(1)} & \cdots &\bm{\mathit{\psi}}_{\omega_{N_f}}^{(M)}\\
  \vert & &\vert
  \end{matrix}}_{\omega_{N_f} }}
  \right]. \label{V}
\end{equation}\\ \\
Here, $\bm{\mathit{\psi}}_{\omega_k}^{(i)}$ are the $i$th SPOD modes of frequency $\omega_k$ and energy $\lambda_{\omega_k}^{(i)}$.
Refer to \citet{schmidt2018spectral} and \citet{towne2018spectral} for more details. This rank-$M\times N_f$ (assuming all $\mathit{\psi}$ are linearly independent) SPOD basis contains
\begin{align}
    E_M=\frac{\sum_{i=1}^{M}\sum_{k=1}^{N_f}\lambda_{\omega_k}^{(i)}}{\sum_{i=1}^{N_b}\sum_{k=1}^{N_f}\lambda_{\omega_k}^{(i)}}\cdot100\% \label{E_M}
\end{align}
of the total energy density.

% Consider a set of time-independent modes $\{\vb*{v}_1,\cdots,\vb*{v}_m\}$ with $\vb*{v}_j\in \mathbb{C}^n$ and $m\leq n$ that span an $m-$dimensional subspace $\mathcal{S}$ of the Hilbert space $\mathcal{H}$ of all possible flow solutions equipped with inner product (\ref{i_prod}).
% Further define $\vb*{V}$ as the matrix with vectors $\vb*{v}_j$ as its columns.

Within the subspace $\mathcal{S}$ spanned by the columns of $\vb*{V}$, the vector $\widetilde{\vb*{q}}(t)$ that best (in a least-squared sense) approximates $\vb*{q}'(t)$ in terms of $\|\cdot\|_E$ is found from the oblique projection
\begin{gather}
    \widetilde{\vb*{q}}(t) = \vb*{P} \vb*{q}'(t), \quad\text{where}\label{oblique} \\
    \vb*{P} =\vb*{V}\left(\vb*{V}^* \vb*{W} \vb*{V}\right)^{-1}\vb*{V}^*\vb*{W}\label{P}
\end{gather}
is the oblique projection matrix for non-singular $\vb*{V}^* \vb*{W} \vb*{V}$.
Alternatively, we may directly express $\widetilde{\vb*{q}}(t)$ as a linear combination 
\begin{align}
    \widetilde{\vb*{q}}(t)= \sum_{i=1}^{M}\sum_{k=1}^{N_f} a_k^{(i)}(t) \bm{\mathit{\psi}}_{\omega_k}^{(i)} = \vb*{V}\vb*{a}(t) \label{expand_q}
\end{align}
of the modes $\bm{\mathit{\psi}}_{\omega_k}^{(i)}$ by defining
\begin{align}
      \vb*{a}(t)\equiv \left(\vb*{V}^* \vb*{W} \vb*{V}\right)^{-1}\vb*{V}^*\vb*{W}\vb*{q}'(t) \label{a_def}
\end{align}
as the vector of temporal expansion coefficients $a_k^{(i)}(t)$. The error vector between the true state and its approximation at any given time is defined as
\begin{align}
    \vb*{e}(t)\equiv\vb*{q}'(t)-\widetilde{\vb*{q}}(t)=\left(\vb*{I}-\vb*{P}\right)\vb*{q}'(t).\label{error}
\end{align}
In the following, we will omit the explicit time dependence of $\vb*{q}'$ and $\vb*{a}$ for brevity. Next, we consider the linearized compressible Navier-Stokes equations
under the linear time-invariant (LTI) assumption,
\begin{align}
      \dv{}{t}\vb*{q}' = \vb*{L}_{\bar{\vb*{q}}}\vb*{q}'+\vb*{f}. \label{dqdt_1}
\end{align}
Here, the matrix $\vb*{L}_{\bar{\vb*{q}}}$ is the discretized linearized Navier-Stokes operator. 

% Galerkin-projection-based order reduction is a well-established technique for both linear and nonlinear systems \citep{benner2015survey,rowley2017model}. For incompressible stationary flows, the quadratic nonlinearity of the Navier-Stokes equations takes the form of a bilinear form that is explicitly accounted for by Galerkin-ROMs \citep{noack2003hierarchy}. When $\vb*{\overline{q}}$ is slowly varying in time, cubic nonlinearities result from the interaction between the mean field and the Reynolds stresses. This problem has been studied by \citet{aubry1988dynamics,holmes1997low}. Standard Galerkin ROMs for systems with quadratic nonlinearities can be applied to compressible flows, but additional modeling assumptions have to made to simplify the more complicated nonlinearities \citep{rowley2004model}.
Different from standard Galerkin ROMs, we model linear dynamics and use the forcing as the compensation for the linearization. First, we obtain the forcing 
\begin{equation}
  \begin{aligned}
 \vb*{f}&=\dv{}{t}\vb*{q}' -\vb*{L}_{\bar{\vb*{q}}}\vb*{q}'
\end{aligned}  
\end{equation}
as the offset between the linear approximation and the true state. This procedure guarantees that the forcing is consistent with the discretization. 
The resulting Galerkin ROM of the LTI system
\begin{align}
    \ \dv{}{t} \vb*{a}=\left(\vb*{V}^* \vb*{W} \vb*{V}\right)^{-1} \vb*{V}^* \vb*{W}\left(\vb*{L}_{\bar{\vb*{q}}}\vb*{q}'+\vb*{f}\right)=\vb*{L}_\text{Gal}\vb*{a}+\vb*{b} \label{L_ROM}
\end{align}
governs the evolution of the forced state. Both the state and the forcing are expressed in terms of their expansion coefficients, $\vb*{a}$ and $\vb*{b}$, respectively.
In equation (\ref{L_ROM}), 
\begin{align}
    \vb*{L}_\text{Gal}= \left(\vb*{V}^*\vb*{W}\vb*{V}\right)^{-1}\vb*{V}^*\vb*{W}\vb*{L}_{\bar{\vb*{q}}}\vb*{V}  \label{L_red}
\end{align}
is the system dynamics matrix, and 
\begin{align}
    \vb*{b}\equiv \left(\vb*{V}^*\vb*{W}\vb*{V}\right)^{-1}\vb*{V}^*\vb*{W}\vb*{f}=\dv{}{t}\vb*{a} -\vb*{L}_\text{Gal}\vb*{a} \label{b}
\end{align}
the vector of expansion coefficients of the forcing field $\vb*{f}$. Equation (\ref{L_ROM}) is an $m$-dimensional first-order differential equation. If a particular trajectory, for example starting from $\vb*{q}'(0)$ is of interest, the corresponding initial condition can be found as $\vb*{a}(0)=\left(\vb*{V}^* \vb*{W} \vb*{V}\right)^{-1}\vb*{V}^*\vb*{W}\vb*{q}'(0)$.
If $\vb*{a}$ is statistically stationary, then 
\begin{equation}\label{b_bar}
  \begin{aligned}
 \overline{\vb*{b}}&=\dv{}{t}\overline{\vb*{a}} -\vb*{L}_\text{Gal}\overline{\vb*{a}}=0
\end{aligned}  
\end{equation}
implies that $\vb*{b}$ has zero mean. This property is important in the context of inverse stochastic models, described next.

\subsection{Inverse stochastic models}
Inverse stochastic models are data-driven models that do not require knowledge of the linear operator $\vb*{L}_\text{Gal}$. Instead, they approximate its action on the model coefficients from the data, most commonly using a least squares approximation and model the stochasticity of the original process as additive noise.

\subsubsection{Linear inverse model (LIM)}
The simplest inverse stochastic model is the linear inverse model (LIM) 
\begin{align} \label{LIM}
 \dv{}{ t}\vb*{a} &=\widetilde{\vb*{T}} \vb*{a} +  \vb*{w},
\end{align}
proposed by \citet{penland1989random,penland1996stochastic}. It assumes a deterministic linear operator that is forced by white noise, $\vb*{w}$. The operator itself is approximated from the data in a least squares sense as
\begin{align}\label{T_tilde}
   \widetilde{\vb*{T}}=\arg\min_{\widetilde{\vb*{T}}}\sum_{j=1}^N\left( \norm{\dv{}{t}{\vb*{a}}^{(j)}-\widetilde{\vb*{T}}\vb*{a}^{(j)}}^2\right).
\end{align}
Note that knowledge of $\vb*{L}_\text{Gal}$ is not required for this procedure. The underlying assumption of the LIM is that the residue of the linear regression, 
\begin{align}\label{rLIM}
     \vb*{r}= \dv{}{t}{\vb*{a}}-\widetilde{\vb*{T}} \vb*{a},
\end{align}
can be approximated as white noise. If the governing linear operator $\vb*{L}_\text{Gal}$ is known, we may express $\widetilde{\vb*{T}}$ as the sum of two matrices,
\begin{align}\label{Ttilde}
    \widetilde{\vb*{T}}&=\vb*{L}_\text{Gal}+\vb*{T}.
\end{align}
Combining equations (\ref{L_ROM}), (\ref{LIM}) and (\ref{Ttilde}) yields the relation 
\begin{align}
    {\vb*{b}}  &=\vb*{T}\vb*{a} + \vb*{w} \label{LIM_0}
\end{align}
between the forcing coefficients, $\vb*{b}$, and the model coefficients, $\vb*{a}$ in terms of $\vb*{T}$. If the feedback matrix $\vb*{T}$ is known, a random realization of the forcing coefficients can be generated to drive the stochastic model. Observe that realizations of $\vb*{b}$ are generated from white noise forcing. This implies that $\vb*{T}$ accounts for the correlations between the state and forcing.
 
%Later, as improvement of LIMs, \citet{kondrashov2005hierarchy,kravtsov2005multilevel} introduced the polynomial inverse models (PIMs) to estimate the nonlinear dynamics by substituting the linear portion in LIMs with a polynomial, in particular, a quadratic dependence.
%The quadratic inverse models yield the general form of 
% \begin{align}
%     \vb*{a} \dd t=\left(\vb*{a}\cdot \mathcal{N}\cdot \vb*{a}+
%     \widetilde{\vb*{T}}\vb*{a} +\vb*{c}\right)\dd t+\dd \vb*{r}^{(0)},
% \end{align}
% where $\mathcal{N}$ is a third-order Tensor, and $\vb*{c}$ is a constant vector. To obtain the parameters of quadratic inverse models, quadratic regression is perform in the form of
% \begin{align}
%     \left[\mathcal{N},\widetilde{\vb*{T}},\vb*{c}\right]=\arg\min_{\mathcal{N},\widetilde{\vb*{T}},\vb*{c}}\sum_{j=1}^N\left( \|\dot{\vb*{a}}^{(j)}-\left(\vb*{a}^{(j)}\cdot \mathcal{N}\cdot \vb*{a}^{(j)}+
%     \widetilde{\vb*{T}}\vb*{a}^{(j)}+\vb*{c}\right)\|^2\right).
% \end{align}

\subsubsection{Linear multi-level regression (MLR) models} \label{MLR_S}

The LIM discussed above assumes that the residue in the model's definition, equation (\ref{rLIM}), is white in time. In practice, however, $\vb*{r}$ is computed from the data using equations (\ref{T_tilde}) and (\ref{rLIM}). Hence, there is no guarantee that the white-noise assumption holds for any given nonlinear process. To address this problem, \citet{kondrashov2005hierarchy,kravtsov2005multilevel} 
introduced the so-called linear multi-level regression (MLR) model. The idea behind the MLR model is to inflate the original model by a hierarchy of additional levels. Each level describes the dynamics of the residue of the previous level and is found by linear regression. Closure of the model is archived once the white-noise assumption holds. Denoting by $\vb*{r}_{1}$ the first-order residue, i.e., the residue of the original model obtained using equation (\ref{rLIM}), the first-level model takes the form $\dv{}{t}{\vb*{a}} = 
    \widetilde{\vb*{T}} \vb*{a} +\vb*{r}_{1}$. The second-level system is then obtained by inflating the state vector by the first-order residue, $\left[ \begin{smallmatrix}  \vb*{a} \\ \vb*{r}_{1} \end{smallmatrix}\right]$, and so on. Assuming that the first $(L-1)$ residues are differentiable, the linear MRL model takes the form
\begin{equation}
  \begin{aligned}
    \text{Level 1:}\quad &\dv{}{t}{\vb*{a}} &=& 
    \widetilde{\vb*{T}} \vb*{a} +\vb*{r}_{1} ,\\
    \text{Level 2:}\quad &\dv{}{t}{\vb*{r}}_{1} &=& \vb*{M}_1  \mqty[\vb*{a}\\ \vb*{r}_{1}] +\vb*{r}_{2} ,\\
    & \vdots \\
   \text{Level $L$:}\quad &\dv{}{t}{\vb*{r}}_{L} &=& \vb*{M}_{L} \mqty[\vb*{a}\\ \vb*{r}_{1}\\
    \vdots \\
    \vb*{r}_{L}] +  {\vb*{w}} .
    \label{MLR}
\end{aligned}  
\end{equation}
Here, $\vb*{M}_{l}$ is the system matrix of size $MN_f\times (l+1)MN_f$, and $\vb*{r}_{l}$ the $l$-th level residue. Alternatively, equation (\ref{MLR}) can be written in the matrix form as 
\begin{align}\label{MLR_M}
    \dv{}{t}\mqty[\vb*{a}\\ \vb*{r}_{1}\\
    \vdots \\
     \vb*{r}_{L-1}\\
    \vb*{r}_{L}]=
    \left[
\begin{array}{c c c c c r}
    \begin{matrix}
    \begin{matrix}
    \begin{matrix} 
    \widetilde{\vb*{T}} & \vb*{I} 
    \end{matrix} &\vb*{0} &  \cdots 
    \end{matrix} & \vb*{0} \\%\hdashline[2pt/2pt]
    \begin{matrix}
    \mathrel{\vcenter{\hbox{\rule{0.4cm}{0.5pt}}}} \vb*{M}_1 \mathrel{\vcenter{\hbox{\rule{0.4cm}{0.5pt}}}}
    &\vb*{I}  & \ddots 
    \end{matrix} & \vdots \\%\hdashline[2pt/2pt]
    \begin{matrix}
     \ddots &   &  &\ddots 
     \end{matrix} & \vb*{0} \\%\hdashline[2pt/2pt]
     \mathrel{\vcenter{\hbox{\rule{1.1cm}{0.5pt}}}} \vb*{M}_{L-1}   \mathrel{\vcenter{\hbox{\rule{1.1cm}{0.5pt}}}}    & \vb*{I}
    \end{matrix}  \\%\hdashline[2pt/2pt]
    \mathrel{\vcenter{\hbox{\rule{1.2cm}{0.5pt}}}} \vb*{M}_{L} \mathrel{\vcenter{\hbox{\rule{1.2cm}{0.5pt}}}}
\end{array}
\right]
    \mqty[\vb*{a}\\ \vb*{r}_{1}\\
    \vdots \\
    \vb*{r}_{L-1}\\
    \vb*{r}_{L}]+\mqty[\vb*{0}\\ \vdots \\ \vdots \\ \vb*{0} \\ \vb*{w} ]. 
\end{align}
Equation (\ref{MLR}) reduces to the classical LIM if $\vb*{r}_1$ is white-in-time to start with.

\section{The stochastic two-level SPOD-Galerkin model} \label{SPOD-Galerkin ROM}

The underlying idea of the proposed approach is to model a statistically stationary flow as the superposition of large-scale coherent structures that evolve on the mean flow under the influence of stochasticity.
This idea is reflected, for example, in the well-known triple-decomposition introduced by \citet{hussain1970mechanics}. SPOD modes optimally represent the second-order space-time statistics of the stationary flow field. We choose SPOD modes as the basis of our model for this reason. The triple-decomposition further requires the `background turbulence' to be stochastic in nature. To accomplish this goal, we employ a multi-level linear regression model, as introduced in \S \ref{MLR_S}. 

In the following, we will demonstrate that the standard Galerkin-projection approach shown in \S \ref{galerkin} is particularly well-suited for this purpose as it requires only one additional level for closure. \S \ref{sec:2lvlROM} introduces the two-level SPOD-Galerkin model and discusses its closure via residue modeling. The final, stochastically driven two-level SPOD-Galerkin model is summarized in \S \ref{sec:stoch2lvlROM}. Finally, an uncertainty quantification study of the new model is conducted in \S \ref{sec:UQ}.
 
\subsection{Two-level SPOD-Galerkin model}\label{sec:2lvlROM}

To obtain the most compact representation of the dynamics, it is desirable to truncate the multi-level model, equation (\ref{MLR}), at the lowest possible level. Our natural starting point is therefore the two-level model
\begin{align}\label{2LM}
    \text{Level 1:}\quad &\dv{}{t}{\vb*{a}} = 
    \vb*{L}_\text{Gal} \vb*{a} +\vb*{b} ,\\
    \text{Level 2:}\quad &\dv{}{t}{\vb*{b}} = \vb*{M}  \vb*{y} +\vb*{r}, \qquad\text{where}\qquad \vb*{y}\equiv\mqty[\vb*{a}\\ \vb*{b}].\label{lvl1}
\end{align}  
Motivated by the goal of creating a physics-based model that relies on regression for closure only, we deviate from the standard linear inverse modeling approach and retain the physics-based operator $\vb*{L}_\text{Gal}$ at first level. Consistent with equation (\ref{L_ROM}), the residue at first level is identified as the forcing vector $\vb*{b}$, obtained from equation (\ref{b}).

%Generally, the operator $\vb*{L}_\text{Gal}$ is not stable, i.e. $\max{\{\mathrm{Re}(\lambda(\vb*{L}_\text{Gal}))\}>0}$, as a consequence of the spatial discretization of the flow field. 
%Therefore, an appropriate control input $\vb*{b}$ is necessary to stabilize the Galerkin ROM without sacrificing accuracy. We then introduce the first level to approximate the dynamics of $\vb*{b}$

Following the standard linear inverse modeling paradigm, we seek closure at the second level by solving the least squares problem 
\begin{align}
   \vb*{M}=\arg\min_{\vb*{M}}\sum_{j=1}^N\left( \norm{\dv{}{t}{\vb*{b}}^{(j)}-\vb*{M}\vb*{y}^{(j)}}^2\right)
\end{align}
to obtain $\vb*{M}$. This least squares problem is formally equivalent to a linear system problem  
\begin{align}
    \dv{}{t}{\vb*{B}}=\vb*{M} \vb*{Y}, \label{M0}
\end{align}
where $\vb*{Y}=\left[\vb*{y}^{(1)},\vb*{y}^{(2)},\cdots,\vb*{y}^{(N)}\right]$ and $\vb*{B}=\left[\vb*{b}^{(1)},\vb*{b}^{(2)}, \cdots, \vb*{b}^{(N)}\right]$ are the matrices of extended vectors and forcing vectors, respectively. The solution of equation (\ref{M0}) is
\begin{align}
    \vb*{M}=\left(\dv{}{t}{\vb*{B}}\right) \vb*{Y}^{+},\label{Lf}
\end{align}
where $\vb*{Y}^+$ denotes the pseudo-inverse of the (possibly singular) extended state matrix. Once $\vb*{M}$ is known, we can obtain the second-level residue $\vb*{r}$ from equation (\ref{lvl1}) as
\begin{equation}\label{residue_0}
  \begin{aligned}
  \vb*{r}&= \dv{}{t}{\vb*{b}} - \vb*{M} \vb*{y}.
\end{aligned}  
\end{equation}
Form equation (\ref{b_bar}), we may deduce that the second-level residue $\vb*{r}$ has zero mean, 
\begin{equation*}
  \begin{aligned}
  \overline{\vb*{r}}&=\overline{\dv{}{t}\vb*{b}}- \vb*{M} \overline{\vb*{y}}=0.
\end{aligned}  
\end{equation*}
The residue $\vb*{r}$ will be approximated as white noise that can be correlated in the subspace, described next. Once $\vb*{M}$ and $\vb*{r}$ are determined, the final two-level model is assembled as
\begin{align}
    \dv{}{t}{\vb*{y}}=\underbrace{
    \left[
\begin{array}{c r}
    \begin{matrix} \vb*{L}_\text{Gal} & \vb*{I} \end{matrix} \\
     \mathrel{\vcenter{\hbox{\rule{0.3cm}{0.5pt}}}}\vb*{M} \mathrel{\vcenter{\hbox{\rule{0.3cm}{0.5pt}}}}
\end{array}
\right]
    }_{\vb*{L}_\text{2-lvl}}\vb*{y}+\mqty[\vb*{0}\\ \vb*{r} ]. \label{dydt_0}
\end{align}
in the form of equation (\ref{MLR_M}). Equation (\ref{dydt_0}) is a forced first-order linear time-invariant system for the extended state vector. The final system dynamic matrix $\vb*{L}_\text{2-lvl}$ contains both the operator governing the linear evolution of large-scale coherent structures---represented by the basis vectors---about the mean flow, $\vb*{L}_\text{Gal}$, and correlation information between the state and the forcing that is learned from the data in $\vb*{M}$. In particular, the left square matrix $\vb*{M}_{ab}$ and the right square matrix $\vb*{M}_{bb}$ that constitute 
\begin{equation}
\vb*{M}=\mqty[\vb*{M}_{ab} & \vb*{M}_{bb}]   \label{M}
\end{equation}
contain correlations between the expansion coefficients and the forcing, and the correlation between the forcing components, respectively. If the residual is computed from the data by means of equation (\ref{residue_0}), the proposed model, equation (\ref{dydt_0}), accurately reproduces the original flow field over all times. The remaining task is to find a suitable model for the residue $\vb*{r}$.

\subsubsection{Residue modelling} \label{residue}
In particular, we seek a stochastic forcing model $\widetilde{\vb*{r}}$ that has the same second-order statistics as the residue $\vb*{r}$. The autocorrelation function that describes the second-order statistics of the residue $\vb*{r}$ at zero time-lag is readily obtained from the data as 
$
    \vb*{R}_{\vb*{r r}}=\overline{\vb*{r}\vb*{r}^*}.
$
Assuming that the residue $\vb*{r}$ is normally distributed in time with 
\begin{equation}\label{rdt_dist}
\vb*{r}\Delta t \sim \mathcal{N} (0,\vb*{R}_{\vb*{r r}}\cdot(\Delta t)^2),
\end{equation}
we may randomly generate $\widetilde{\vb*{r}}$ from Gaussian white noise $\vb*{w} \sim \mathcal{N}(0,\vb*{I})$ as
\begin{align}\label{r_tilde}
    \widetilde{\vb*{r}} = \vb*{G w},
\end{align}
where $\vb*{G}$ is the unknown input distribution matrix. The role of $\vb*{G}$ in our model is to color randomly generated white noise with the statistics of the residual. \emph{Vice versa}, $\vb*{G}^{-1}$ can be interpreted as a whitening filter \citep{van1968detection}. Gaussian white noise can be generated, by sampling from a Wiener process $\vb*{\xi}$, as
\begin{align}
    \vb*{w}\equiv\dv{\vb*{\xi}}{t}=\lim_{\Delta t\to 0}\frac{\vb*{\xi}(t+\Delta t)-\vb*{\xi}(t)}{\Delta t}.
\end{align}
For any time increment $\Delta t$, $\Delta \vb*{\xi}$ is normally distributed with zero mean and covariance matrix $\vb*{I}\Delta t$. We may now express equation (\ref{r_tilde}) in terms of $\Delta \vb*{\xi}$ as
\begin{align}\label{eqn:rtildedt}
   \widetilde{ \vb*{r}}\Delta t= \vb*{G}\Delta \vb*{\xi},
\end{align}
where
\begin{align} \label{eqn:Gdxi}
    \vb*{G}\Delta \vb*{\xi} \sim \mathcal{N} (0, \vb*{G G^*} \Delta t)
\end{align}
follows from the properties of the Wiener process. Comparing equations (\ref{eqn:Gdxi}) and  (\ref{rdt_dist}) allows us to relate the unknown input distribution matrix, $\vb*{G}$, to the known covariance matrix of the residue, $\vb*{R}_{\vb*{r r}}$, as
\begin{align}
    \vb*{R}_{\vb*{r r}}\Delta t= \vb*{G G^*}.  \label{G}
\end{align}
% which yields the distribution
% \begin{align}
%     \vb*{r}\Delta t \sim \mathcal{N} (0, \vb*{G G^*} \Delta t). 
% \end{align}
Based on the form of equation (\ref{G}), we obtain $\vb*{G}$ from the Cholesky decomposition of $\vb*{R}_{\vb*{r r}}\Delta t$. Owing to this procedure of computation, $\vb*{G}^{-1}$ may be referred to as a Cholesky whitening filter.

% The residue $\vb*{r}$ can then be approximated as
% \begin{align}
%     \vb*{r} \approx \vb*{G w},
% \end{align}
% where $\vb*{G w}$ is a vector of spatially correlated Gaussian white noise \citep{kondrashov2015data}. 

% We further define a vector of Gaussian, white-noise random variables $\vb*{w}$ from the Wiener process by 
% \begin{align}
%     \vb*{w}\dd t= \dd \vb*{\xi},
% \end{align}
% which satisfies the distribution $\vb*{w} \sim \mathcal{N}(0,\vb*{I})$.

% To hold the correlation in the subspace, we approximate the time increment $\vb*{r} \dd t$ as correlated multivariate additive noise
% \begin{align}
%     \vb*{r}\dd t \approx \vb*{G} \dd \vb*{\xi}, \label{Gxi}
% \end{align}
% where $\vb*{G}$ is an input distribution matrix, and $\vb*{\xi}$ is the vector of Gaussian Wiener processes. 

% We assume the residue $\vb*{r}$ obeys the distribution 
% \begin{align}
%     \vb*{r}\sqrt{\Delta t} \sim \mathcal{N} (0,\vb*{R}_{\vb*{r r}}\Delta t).  \label{rdt}
% \end{align}

\subsection{Stochastic two-level SPOD-Galerkin model}\label{sec:stoch2lvlROM}
Closure of the model is achieved by introducing the stochastic forcing model $\widetilde{\vb*{r}}$ into equation (\ref{dydt_0}). The resulting stochastic two-level SPOD-Galerkin model takes the form of the stochastic differential equation
\begin{align}
    \mathrm{d}{\vb*{y}} = \vb*{L}_\text{2-lvl} \vb*{y} \,\mathrm{d}t +\int_t^{t+\mathrm{d}t} {\widetilde{\vb*{w}}} \,\mathrm{d}t , \; \text{where}\quad \widetilde{\vb*{w}}\equiv \mqty[\vb*{0} \\\vb*{G w}] \label{sde}
\end{align}
is the process noise. Equation (\ref{sde}) is a stochastic differential equation (SDE) that can be solved numerically by approximating the time derivative using a forward difference as
\begin{align}
      \vb*{y}^{(j+1)} =\vb*{H y}^{(j)} +
      \Delta\widetilde{\vb*{\xi}}^{(j)},\label{sde_d}
\end{align}
where $j$ is the time index, $\Delta t$ the time step, and
\begin{align}
\vb*{H}=\vb*{I}+\vb*{L}_{\text{2-lvl}}\Delta t \quad\text{and}\quad \Delta \widetilde{\vb*{\xi}}\equiv \mqty[\vb*{0} \\\vb*{G } \Delta \vb*{\xi}] \label{H}
\end{align}
are the state transition matrix and the process noise, respectively. By construction, the process noise $\Delta \widetilde{\vb*{\xi}}$ is a zero-mean Gaussian random sequence with covariance matrix
\begin{align}
    \widetilde{\vb*{R}}_{\vb*{r r}} \equiv \overline{\Delta \widetilde{\vb*{\xi}} \Delta \widetilde{\vb*{\xi}}^*}=\mqty[ \vb*{0} & \vb*{0}\\
    \vb*{0}& \vb*{R}_{\vb*{r r}}\Delta t].
\end{align}
This completes the model. The discrete-in-time algorithm is outlined in the following.

% 	\begin{figure} 
% 		\centering
%         \includegraphics[trim = 10mm 15mm 20mm 10mm, clip, width=1\textwidth]{Schematic.pdf}
%         \caption{Schematic of the stochastic two-level SPOD-Galerkin  model. Variables related to the input data are marked in blue, nonlinear forcing in red, compound state in magenta, and residue in green. The discrete-in-time algorithm is outlined in \S\ref{alg}. } \label{schematic}
% 	\end{figure}
% A schematic of the complete stochastic two-level SPOD-Galerkin model is shown in figure \ref{schematic}. 
% \Red{We first construct the SPOD-Galerkin ROM by performing Galerkin projection on the linearized compressible Navier-Stokes equations based on SPOD modes. The linear two-level regression then models the resulting reduced-order kinematics. The closure of the model is achieved by whitening the error of the linear regression process.
% } 

\subsection{Algorithm: stochastic two-level SPOD-Galerkin}\label{alg}

\noindent {\bf Algorithm} Stochastic two-level SPOD-Galerkin model\\
{\bf Input:} Fluctuating data matrix $\vb*{Q}'=\left[\vb*{q}^{(1)}-\overline{\vb*{q}},\vb*{q}^{(2)}-\overline{\vb*{q}},\cdots,\vb*{q}^{(N)}-\overline{\vb*{q}}\right]$ consisting $N$ snapshots, discretized linearized Navier-Stokes operator $\vb*{L}_{\bar{\vb*{q}}}$, time step $\Delta t$. \\
{\bf Output:} Matrices $\vb*{L}_\text{2-lvl}$ and $\vb*{G}$ of the stochastic two-level SPOD-Galerkin model $\dv{}{t}{\vb*{y}} = \vb*{L}_\text{2-lvl} \vb*{y} + \widetilde{\vb*{w}}$, where $\widetilde{\vb*{w}}= \mqty[\vb*{0} \\\vb*{G w}]$.
\begin{enumerate}
    \item Compute the SPOD of $\vb*{Q}'$ and store the SPOD modes to be retained in the model in the column (basis) matrix $\vb*{V}$.
        \item Determine the expansion coefficients and the Galerkin system dynamics matrix as 
        \begin{align*}
            \vb*{A} = \left(\vb*{V}^* \vb*{W} \vb*{V}\right)^{-1}\vb*{V}^*\vb*{W}\vb*{Q}' \quad \text{and} \quad \vb*{L}_{\text{Gal}} = \left(\vb*{V}^* \vb*{W} \vb*{V}\right)^{-1}\vb*{V}^*\vb*{W}\vb*{L}_{\bar{\vb*{q}}}\vb*{V},
        \end{align*}
        respectively.
        \item Following the equation (\ref{b}), calculate the forcing coefficients as 
        \begin{align*}
             \vb*{B} = \frac{\vb*{A}_2^{N}-\vb*{A}_1^{N-1}}{\Delta t}- \vb*{L}_{\text{Gal}}\vb*{A}_{1}^{N-1}.
        \end{align*}
        where $\vb*{A}_2^{N}=\left[\vb*{a}^{(2)},\vb*{a}^{(3)},\cdots,\vb*{a}^{(N)}\right]$ and $\vb*{A}_1^{N-1}=\left[\vb*{a}^{(1)},\vb*{a}^{(2)},\cdots,\vb*{a}^{(N-1)}\right]$.
        \item Let $\vb*{Y}\leftarrow\mqty[\vb*{A}_1^{N-1}\\ \vb*{B}]$. Solving the linear system problem in the equation (\ref{Lf}) yields
        \begin{align*}
            \vb*{M} = \frac{\vb*{B}_2^{N-1}-\vb*{B}_1^{N-2}}{\Delta t} \vb*{Y}^*\left(\vb*{Y Y}^*\right)^{-1}.
        \end{align*}
        \item Assemble the matrices to obtain  
        \begin{align*}
            \vb*{L}_{\text{2--lvl}} \leftarrow     \left[
\begin{array}{c r}
    \begin{matrix} \vb*{L}_\text{Gal} & \vb*{I} \end{matrix} \\
     \mathrel{\vcenter{\hbox{\rule{0.3cm}{0.5pt}}}}\vb*{M} \mathrel{\vcenter{\hbox{\rule{0.3cm}{0.5pt}}}}
\end{array}
\right].
        \end{align*}.
        \item Calculate the linear regression residue as
                \begin{align*}
             \vb*{R} = \frac{\vb*{B}_2^{N-1}-\vb*{B}_1^{N-2}}{\Delta t}- \vb*{M}\vb*{Y}_1^{N-2}.
        \end{align*}
        \item Determine the matrix $\vb*{G}$ by
        solving the Cholesky decomposition
        \begin{align*}
            \vb*{G G}^* = \frac{\Delta t}{N-2}\vb*{R R}^*.
        \end{align*}
\end{enumerate}

The resulting stochastic two-level SPOD-Galerkin model is propagated in time as
\begin{align*}
    \vb*{y}^{(j+1)} =\left[\vb*{I}+\vb*{L}_{\text{2-lvl}}\Delta t\right]\vb*{y}^{(j)} +
      \mqty[\vb*{0} \\\vb*{G } \sqrt{\Delta t} \vb*{w}],
\end{align*}
where $j$ is the time index, see equations (\ref{sde})-(\ref{H}).  The zero mean, unit variance Gaussian white noise $\vb*{w}$ which drives the system is obtained from a random number generator. 
	
\subsection{Uncertainty quantification and spectral analysis}\label{sec:UQ}

% [Specifically, when $\dv{}{t}
%  {\vb*{y}}$ is discretized by finite difference method and the residue $\vb*{r}$ is white noise, the equation (\ref{dydt_0}) is analogous to vector autoregressive (VAR) models.]

Next, we leverage the resemblance of equation (\ref{sde_d}) to first-order vector autoregression (VAR) processes to conduct an uncertainty quantification analysis of the model. Following \citet{stengel1986stochastic}, we analyze the propagation of the uncertainty 
in terms of the expected value of the extended state vector, $\langle \vb*{y}^{(j)}\rangle$, where $\expval{\cdot}$ denotes the average over a large number of realizations of $\Delta\widetilde{\vb*{\xi}}^{(j)}$.  Since the forcing has zero-mean, the equation 
\begin{align}
   \langle \vb*{y}^{(j)}\rangle=\langle\vb*{H y}^{(j-1)} +\Delta \widetilde{\vb*{\xi}}^{(j-1)}\rangle =\vb*{H}\langle \vb*{y}^{(j-1)}\rangle \label{mj}
\end{align}
shows that $\vb*{H}$ functions as the propagator of the expected value of the state. If $\vb*{H}$ is stable with $\max{\{|\lambda(\vb*{H})|\}}<1$, then the average of $\vb*{y}^{(j)}$ at large time, denoted by $ \lim_{j\to \infty} \langle \vb*{y}^{(j)}\rangle$, is zero. Denoting by 
\begin{equation}
    \vb*{y}'^{(j)} = \vb*{y}^{(j)} - \expval{\vb*{y}^{(j)}}\label{yp}
\end{equation}
the fluctuation of the state, we may compute the auto-covariance matrix $\vb*{P}^{(j)}$ of the state as
\begin{eqnarray}
    \vb*{P}^{(j)}&=&\expval{\vb*{y}'^{(j)}\vb*{y}'^{(j)^*}}\nonumber \\
    &=&\vb*{H P}^{(j-1)} \vb*{ H}^*+\widetilde{\vb*{R}}_{\vb*{r r}}. \label{Pj}
\end{eqnarray}
Equation (\ref{Pj}) is the propagation equation for auto-covariance of the state, which is readily obtained by combining equations (\ref{sde_d}), (\ref{mj}), and (\ref{yp}). Here, we made use of the fact that the fluctuating state is uncorrelated to the process noise, that is, $ \langle\vb*{y}'^{(j-1)} \Delta \widetilde{\vb*{\xi}}^{(j-1)^*}\rangle=0$. The reason is that the stochastic component of $\vb*{y}'^{(j-1)}$ is computed from the evaluation of the stochastic process at the previous time instant, $\Delta\widetilde{\vb*{\xi}}^{(j-2)^*}$, see equation (\ref{sde_d}).

If the true state $\vb*{y}^{(1)}$ is used as the initial condition, then $\vb*{P}^{(0)}=0$ from equation \ref{Pj}, and it can be shown that the matrix sequence $\{\vb*{P}^{(j)}\}$ converges to the true covariance
\begin{align}
    \vb*{P}=\lim_{j\to \infty}\vb*{P}^{(j)}=\sum_{n=0}^{\infty}\vb*{H}^n\widetilde{\vb*{R}}_{\vb*{r r}}\left(\vb*{H}^*\right)^n,
\end{align}
which solves the discrete-time Lyapunov equation 
\begin{align}
    \vb*{H P H}^*-\vb*{P}+\widetilde{\vb*{R}}_{\vb*{r r}}=0.
\end{align}
Since $\widetilde{\vb*{R}}_{\vb*{r r}}$ is hermitian and positive-definite by construction, 
the existence and uniqueness of the solution $\vb*{P}$ are guaranteed by the Lyapunov theorem if $\vb*{H}$ is stable. 
Since the process noise $\Delta \widetilde{\vb*{\xi}}^{(j)}$ is Gaussian, the distribution of the realizations of the state, $\vb*{y}$, is also Gaussian for the LTI system at hand. 
Therefore, the $95\%$ confidence interval of $\vb*{y}$ at $t_j$ can be determined as 
\begin{equation}\label{conf}
\mathrm{CI}=\left(\langle \vb*{y}^{(j)}\rangle- 2 \sqrt{\text{diag}(\vb*{P}_j)},\;\langle \vb*{y}^{(j)}\rangle+ 2 \sqrt{\text{diag}(\vb*{P}_j)}\right).    
\end{equation}
In the limit of large times with $j\rightarrow\infty$, this interval converges to the bounded interval $\left(-2 \sqrt{\text{diag}(\vb*{P})}, 2 \sqrt{\text{diag}(\vb*{P})}\right)$. Under the assumption that the realizations generated by the discrete SDE, equation (\ref{sde_d}), are weakly stationary with zero mean and covariance matrix $\vb*{P}$, the time-lagged autocorrelation function can be determined analytically in terms of the state transition matrix as
\begin{align}
    \vb*{R}_{\vb*{y y} }(n)\equiv\langle\vb*{y}^{(j)}\vb*{y}^{(j+n)^*} \rangle=\vb*{P}\left(\vb*{H}^*\right)^{|n|}.
\end{align}
By way of the Wiener-Khinchin theorem, the analytical expression for the spectral density function 
\begin{equation}
\begin{aligned}
    \vb*{S}_{\vb*{y y}}(\omega)&=\frac{1}{\sqrt{2 \pi}} \sum_{n=-\infty}^{\infty}\vb*{R}_{\vb*{y y}}(n) \mathrm{e}^{-\mathrm{i}\omega n}\\
    &=\frac{1}{\sqrt{2 \pi}} \vb*{P}\left[\left(\vb*{I}-\mathrm{e}^{-\mathrm{i}\omega}\vb*{H}^*\right)^{-1}+\left(\vb*{I}-\mathrm{e}^{\mathrm{i}\omega}\vb*{H}^*\right)^{-1}-\vb*{I}\right].
\end{aligned}
\end{equation}
is found by means of the discrete-time Fourier transform. The diagonal of $\vb*{S}_{\vb*{y y}}(\omega)$ then contains the power spectral density of $\vb*{y}$.

\section{Example of a turbulent jet} \label{example}
Take as an example of a statistically stationary flow a turbulent, iso-thermal jet at Mach number, based on the jet velocity and the far-field speed of sound, of $M=0.9$ and Reynolds number, based on the nozzle diameter and the jet velocity, of $\Re\approx10^6$. 
The state vector
\begin{align}
    \vb*{q}=[\rho,u_x,u_r,u_{\theta},T]^T, \label{q}
\end{align}
comprises the density $\rho$, temperature $T$, and cylindrical velocity components $u_x$, $u_r$ and $u_{\theta}$ in the streamwise, $x$, radial, $r$, and circumferential, $\theta$, directions, respectively. 
For more details, the reader is referred to \citet{bresetal_2018jfm}.
Owing to the rotational symmetry of the jet, we may decompose the data, without loss of generality, into azimuthal Fourier components, $m$.
% 	\begin{figure} 
% 		\centering
%         \includegraphics[trim = 10mm 10mm 10mm 20mm, width=0.95\textwidth]{jet_m0_new.pdf}
%         \caption{Instantaneous flow field of the axisymmetric component of a subsonic jet: ($a$) streamwise
% cross-section along the jet axis; ($b$-$e$) transverse planes at different streamwise locations
% $x$.}\label{jet_m0}
% 	\end{figure}

In the same way, we may decouple the governing equations and construct the linear operator $\vb*{L}_{\bar{\vb*{q}}}$, defined in equation (\ref{dqdt_1}), for each azimuthal wavenumber independently. Upon linearization of the compressible Navier-Stokes equations, we obtain the linearized Navier-Stokes operator $\vb*{L}_{\bar{\vb*{q}}}$ that governs the fluctuating state, $\vb*{{q'}}=[{\rho'},{u'}_x,{u'}_r,u'_\theta,{T'}]^T$, here defined in terms of primitive variables, as
\begin{align}
      \dv{}{t}\vb*{{q'}} = \vb*{L}_{\bar{\vb*{q}}}\vb*{{q'}}. \label{dqdt_prim}
\end{align}
Consequently, $\vb*{\bar{q}}=[\bar{\rho},\bar{u}_x,\bar{u}_r,0,\bar{T}]^T$ is the long-time of the primitive state, whose azimuthal velocity component is zero for the round jet. For more details, the reader is referred to \citet{schmidt2017wavepackets}.
In this example, we construct a stochastic two-level SPOD-Galerkin model for the symmetric component of the jet with $m=0$. We interpolate the data on a $950\times195$ Cartesian mesh that includes the physical domain $x,r \in [0,30] \times [0,6]$.

To quantify the flow energy, we use the compressible energy inner product
\begin{align}
     \left<\vb*{q}_1,\vb*{q}_2\right>_E=\int_{\Omega} \vb*{q}_1^*\mathrm{diag}\left(\frac{\overline{T}}{\gamma \bar{\rho}M^2},\bar{\rho},\bar{\rho},\bar{\rho},\frac{\bar{\rho}}{\gamma(\gamma-1)\overline{T}M^2}\right)\vb*{q}_2 \dd \vb*{x}=\vb*{q}_1^*\vb*{W}\vb*{q}_2, \label{chu_norm}
\end{align}
devised by \citet{chu1965energy}, in equation (\ref{i_prod}). $\vb*{W}$ is the weight matrix containing both the numerical quadrature weights and weights associated with this inner product. The wide range of time and length scales of this fully developed turbulent flow becomes apparent from the velocity field.

\subsection{SPOD of the turbulent jet} \label{SPOD_jet}
We compute the SPOD from the 10,000 snapshots of the turbulent jet by partitioning the data into $N_b=77$ blocks of 256 snapshots with an overlap of $50\%$. 
These spectral estimation parameter are obtained following the best practices outlined in \citet{towne2018spectral,schmidt2020guide}.
Owing to the rotational and temporal symmetry of the jet, it suffices to consider the $N_f=129$ non-negative frequency components. 

	\begin{figure} 
		\centering
        \includegraphics[trim = 0mm 0mm 0mm 0mm, width=0.8\textwidth]{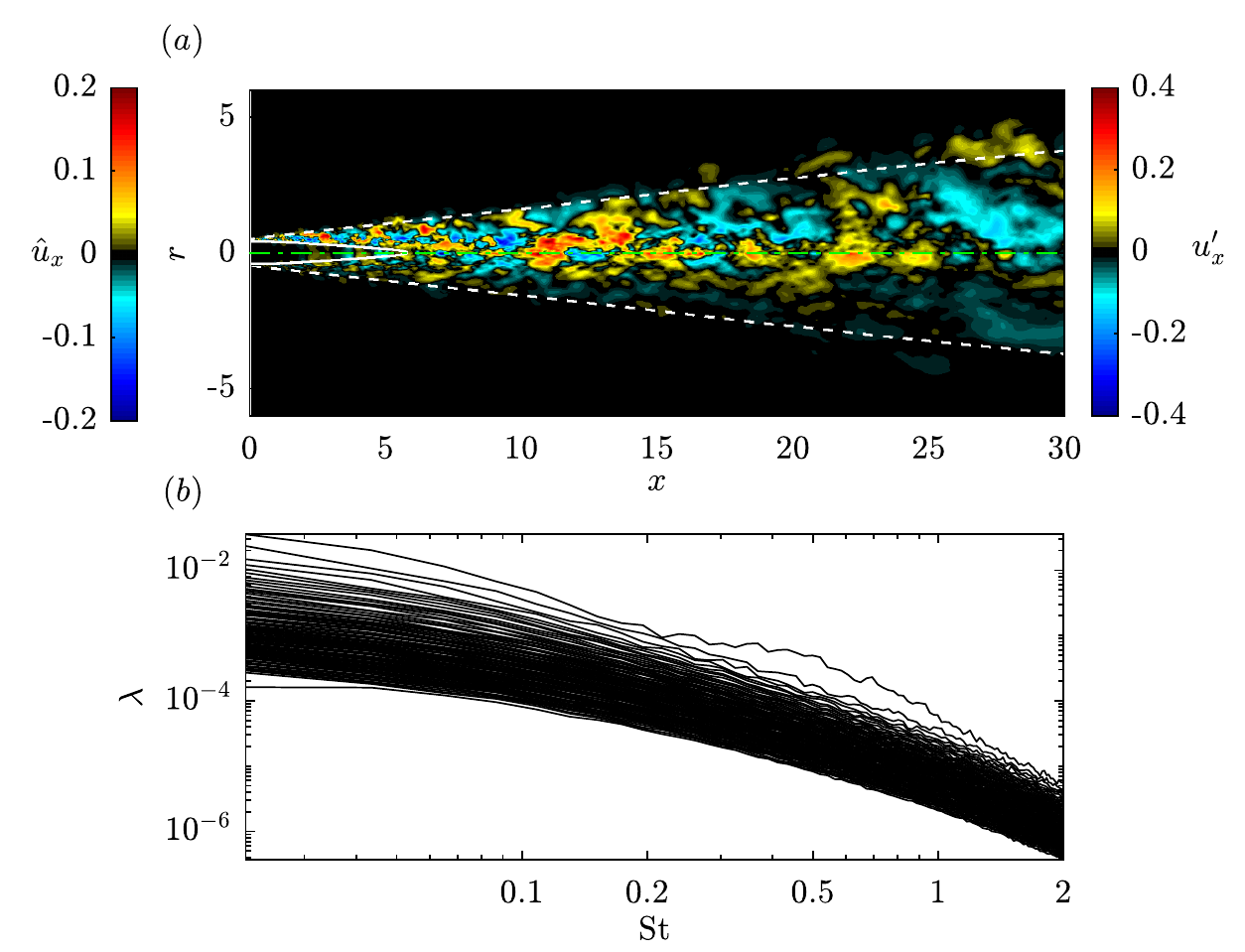}
        \caption{
        Axisymmetric component of a $M=0.9$ transonic jet: (a) streamwise cross-section of the instantaneous streamwise fluctuating velocity (top half) and its symmetric component (bottom half); (b) SPOD eigenvalue spectra. }\label{jet_m0}
	\end{figure}

% 	\begin{figure} 
% 		\centering
%         \includegraphics[width=0.55\textwidth]{eig_spectra_s.pdf}
%         \caption{SPOD eigenvalue spectra for the axisymmetric component of a $M=0.9$ transonic jet. The SPOD eigenvalue problem is solved for each discrete frequency. Each line represents one eigenvalue at each frequency, e.g., the top line represents the leading eigenvalue at each frequency.}\label{spod_spectra}
% 	\end{figure}

Figure \ref{jet_m0}($a$) shows the instantaneous streamwise fluctuating velocity, $u_x'$, and its symmetric component (azimuthal wavenumber $m=0$), $\hat{u}_x$, of the turbulent jet.
Figure \ref{jet_m0}($b$) shows the SPOD eigenvalue spectra of the axisymmetric component. The reader is referred to \citet{schmidt2018spectral} for more details on the physical interpretation of the SPOD.

		\begin{figure} 
		\centering
        \includegraphics[trim = 11mm 14mm 18mm 55mm, clip, width=1\textwidth]{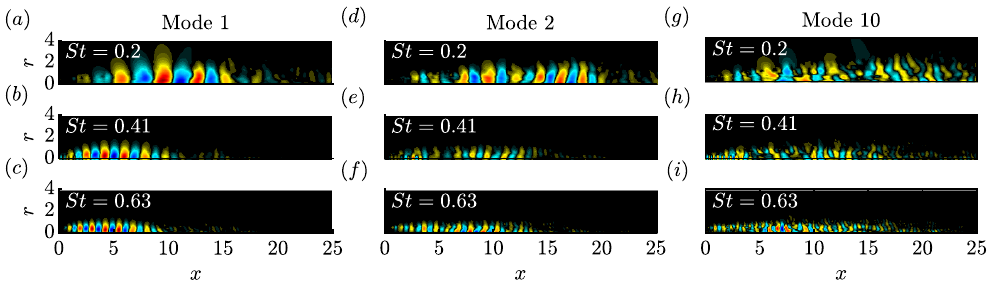}
        \caption{Examples of SPOD modes that form the basis of the model: (a,d,g) $St=0.2$; (b,e,h) $St=0.41$; (c,f,i) $St=0.63$. The normalized pressure components of the 1st, 2nd, 10th modes are 
 shown in $x,r \in [0,25] \times [0,4]$.}\label{spod_modes}
	\end{figure}
Figure \ref{spod_modes} shows the first, the second and the 10th of the SPOD modes that constitute the basis of our models. Three representative frequencies are presented. Large-scale coherent structures associated with Kelvin-Helmholtz instability waves are observed in the leading modes.

% The more chaotic spatial structure of the higher SPOD modes indicates that they are not fully converged. Physically, this lack of convergence reflects the non-low-rank nature of the jet turbulence. It is not a concern in the present context of low-order modeling.
	
\subsection{Subspace modeling}
\begin{figure} 
		\centering
        \includegraphics[trim = 0mm 6mm 0mm 0mm, clip,  width=1 \textwidth]{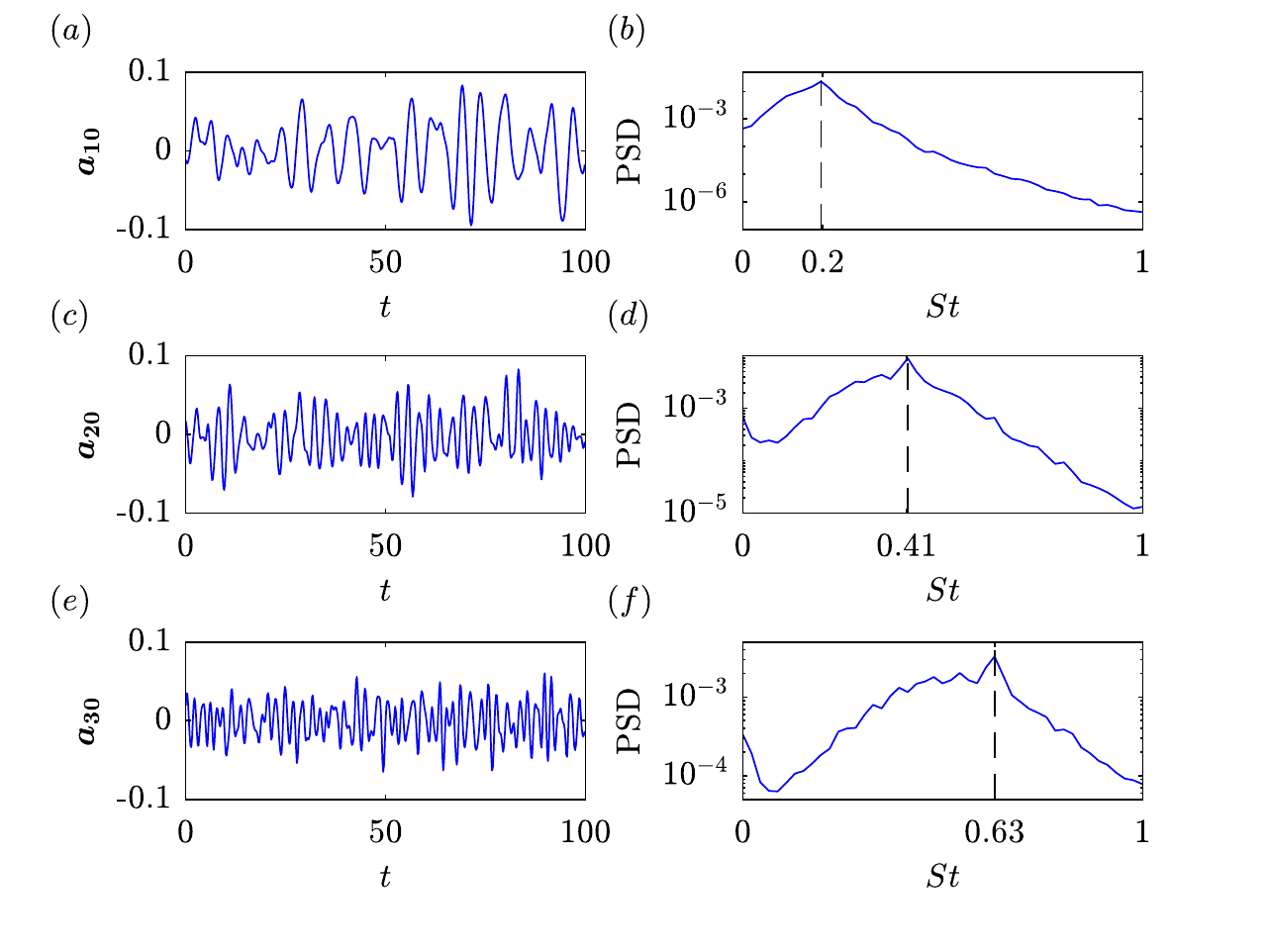}
        \caption{SPOD expansion coefficients $\vb*{a}_i(t)$ for the three representative modes shown in figure \ref{spod_modes}(a-c) as part of the rank $1\times129$ basis: $(a-b)$  $St=0.2$; $(c-d)$ $St=0.41$; $(e-f)$ $St=0.63$. Time trace over 100 time units (left), and PSD (right). The frequency of the SPOD modes are marked by dashed lines in the spectra.}\label{A_q}
	\end{figure}
Figure \ref{A_q} shows the SPOD expansion coefficients, $\vb*{a}_i(t)$, obtained from the oblique projection defined in equation (\ref{a_def}) for the three modes shown in figure \ref{spod_modes}(a-c). Results are shown for the rank $1\times 129$ basis, that is, the basis consisting exclusively of the leading SPOD modes (one mode per frequency). Note again that each SPOD mode is, by construction, associated with a single frequency. To confirm that the oblique projection truthfully represents this property, we compute the periodograms (right column) to reveal the frequency content of the expansion coefficients. It is observed that the PSD indeed peaks at the respective SPOD frequencies. This confirms that the SPOD modes in fact predominantly represent the spectral content they optimally account for by construction. Even though the model does technically not depend on this property, we note that this observation can be interpreted as an \emph{a posteriori} justification for the use of the oblique projection introduced in equation (\ref{a_def}), and therefore also of the use of SPOD modes as a modal basis in the time domain.

% Following (\ref{error}), the error vector in subspace approximation of the flow quantity $\vb*{q}'$ is
% \begin{align}
%   \vb*{e}=\vb*{q}'-\widetilde{\vb*{q}}= (\vb*{I}-\bm{\mathit{\Psi}}\left(\bm{\mathit{\Psi}}^*\vb*{W}\bm{\mathit{\Psi}}\right)^{-1}\bm{\mathit{\Psi}}^*\vb*{W})\vb*{q}'.
% \end{align}	
	\begin{figure} 
		\centering
        \includegraphics[trim = 0mm 20mm 0mm 20mm, clip,width=1\textwidth]{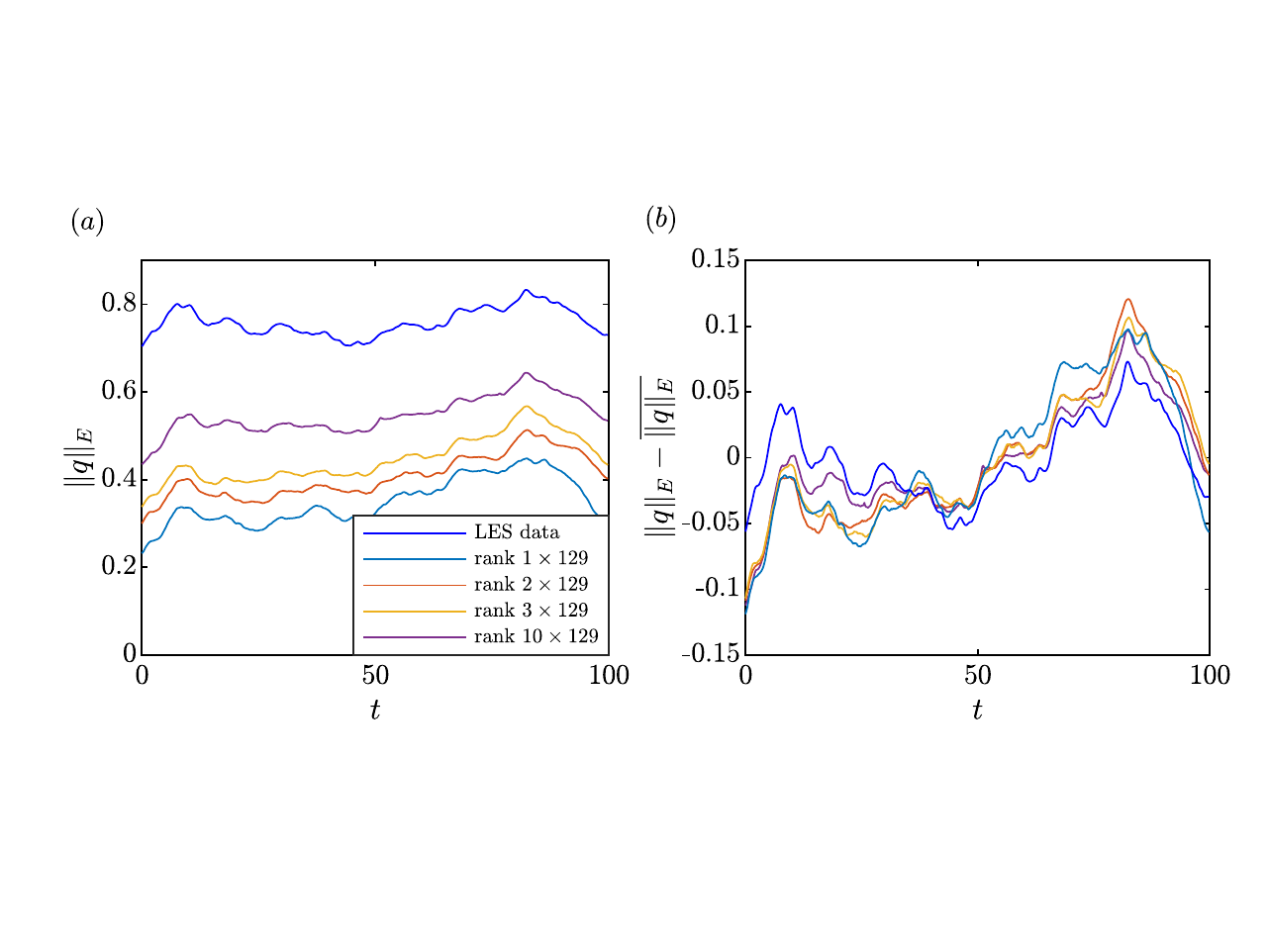}
        \caption{Compressible energy norm in (a), and deviation from the mean energy in (b). The LES data and the low-rank approximations with $13\%$, $21\%$, $27\%$, $52\%$ energy are reported. Part (b) emphasises that the dynamics (deviation from the mean) are well captured even by the lowest-rank approximation.} \label{C_norm}
	\end{figure}
	
Figure \ref{C_norm} (a) compares the compressible energy norm, $\|\vb*{q}(\vb*{x},t)\|_{E}$, of the full LES data to the energy of the partial reconstructions, $\widetilde{\vb*{q}}$, for modal bases of different sizes. It can be observed that all reconstructions follow the dynamics of the LES. The approximately constant offset between the energy of the LES and the different low-rank approximations is similarly reflected in the retained SPOD energy. In particular, the rank $1\times129$, $2\times129$, $3\times129$, and $10\times129$ approximations account for $13\%$, $21\%$, $27\%$, and $52\%$ of the total energy, respectively.  
Figure \ref{C_norm}(b) shows that the offset between the low-rank approximation and the true state is almost constant. This 
implies that the error stems, almost exclusively, from the energy contained in scales that are not important from a dynamics perspective. 

	\begin{figure}
		\centering
        \includegraphics[trim = 5mm 10mm 10mm 0mm, clip, width=0.9\textwidth]{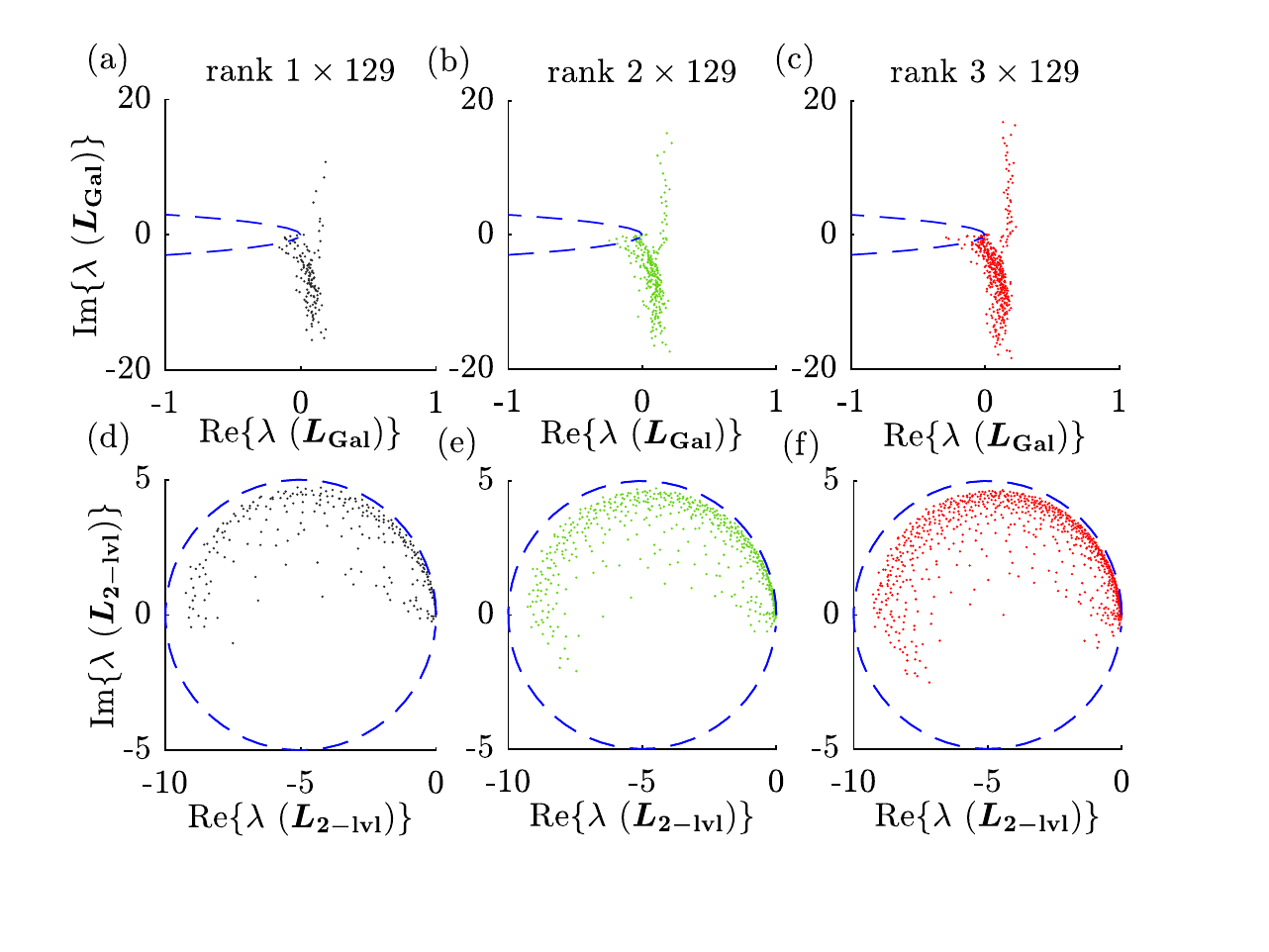}
        \caption{Eigenvalues of the standard SPOD-Galerkin ROM, $\lambda(\vb*{L}_{\text{Gal}})$, and the proposed two-level model, $\lambda(\vb*{L}_{\text{2-lvl}})$: (a,d) rank $1\times 129$ (black); (b,e) $2\times 129$ (green); (c,f) $3\times129$ (red). The blue dashed circle represents the stability region. The eigenvalues outside of the unit disc in (a-c) indicate that Galerkin projection yields an unstable model. On the contrary, the two-level model in (d-f) is stable.} \label{L_ROM_eig}
	\end{figure}

In figure \ref{L_ROM_eig}, the stability of the naive Galerkin ROM of the LTI system, equation (\ref{L_red}), and the two-level model, equation (\ref{dydt_0}), is addressed in terms of the eigenvalue spectra of the corresponding operators, $\vb*{L}_{\text{Gal}}$ and $\vb*{L}_{\text{2-lvl}}$, respectively. The dashed blue line in the spectra corresponds to the disc of radius $\frac{1}{\Delta t}$, centered about $-\frac{1}{\Delta t}$. It demarcates the region of stability; eigenvalues inside the circle are associated with temporal decay, whereas eigenvalues on the outside are associated with exponential amplification. Clearly, the simple Galerkin ROM is unstable, whereas the two-level model is stable. Furthermore, the eigenvalues of the 2-level model are mostly confined to the upper half of the stability region. This behaviour is explained by the symmetry of the SPOD spectrum and the restriction of the modal basis to non-redundant positive frequency content. Figure \ref{L_ROM_eig}(d-f) shows that the eigenvalues of $\vb*{L}_\text{2-lvl}$ remain confined to a specific area within the region of stability when the model rank is increased. These results demonstrate that the transition matrices $\vb*{H}$ computed using equation (\ref{H}) are stable, that is, $\max{\{|\lambda(\vb*{H})|\}}<1$. Hence, when there is no stochastic input to the SDE, equation (\ref{sde_d}), the extended state vector $\vb*{y}$ vanishes as $t\rightarrow \infty$ as its expected value is zero. In accordance with the modeling philosophy, this implies that the appropriate forcing input is necessary to sustain the turbulent flow.

	\begin{figure}
		\centering
        \includegraphics[trim = 0mm 0mm 0mm 0mm, clip, width=0.5\textwidth]{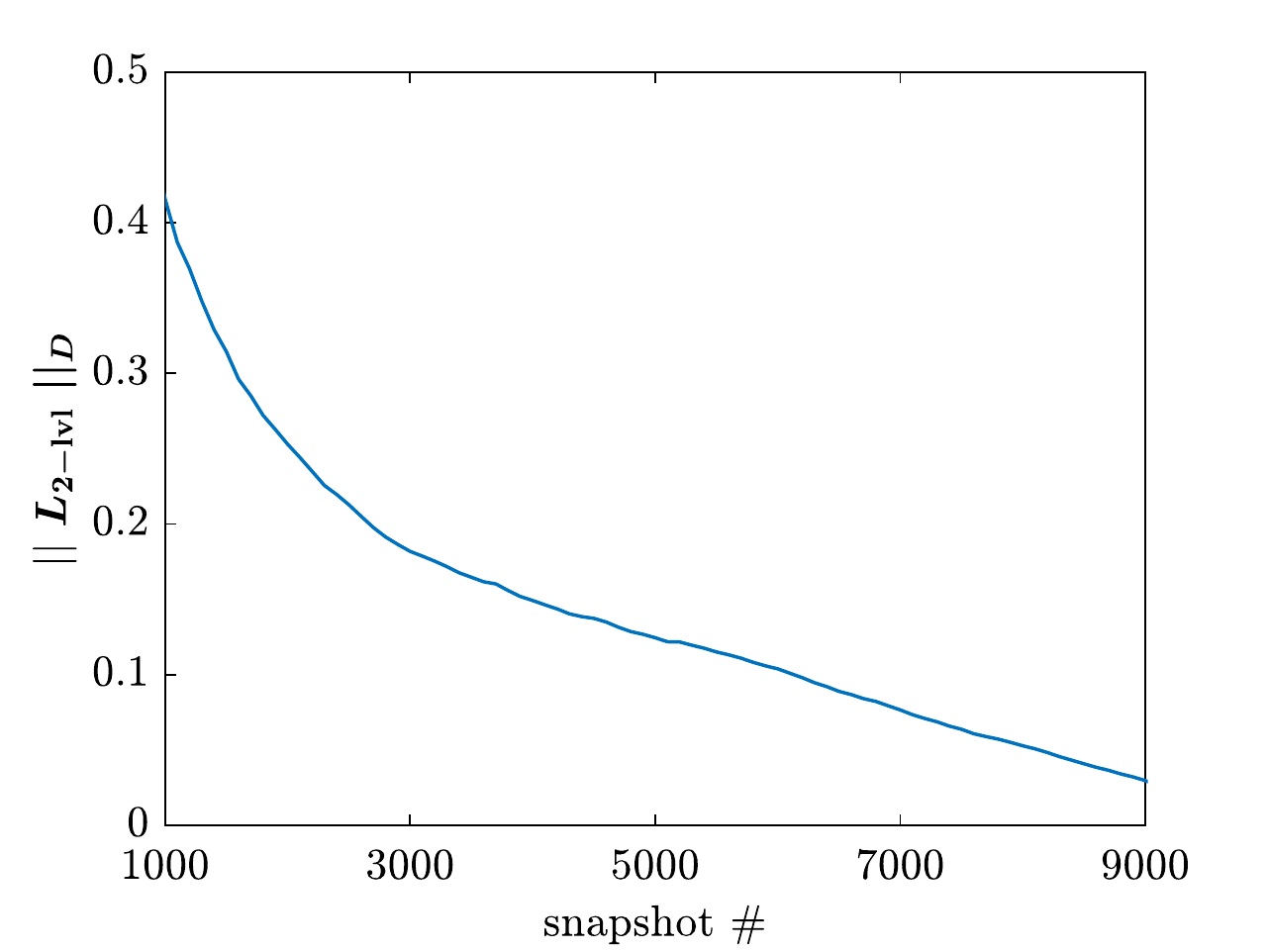}
        \caption{Convergence of the matrix sequence $\{\vb*{L}_\text{2-lvl}^{(n)}\}$. } \label{X_norm}
	\end{figure}

As the turbulence closure relies on linear regression, it has to be shown that $\vb*{L}_{\text{2-lvl}}$ converges as more data is used for its construction. Consider the sequence of matrices, $\{\vb*{L}_\text{2-lvl}^{(n)}\}$, where $n$ is the number of snapshots used in the linear regression, equation (\ref{M0}). We define a normalized norm of these the matrix sequence, $\|\cdot \|_D$, as
\begin{align}
    \|\vb*{L}_\text{2-lvl}\|_D &\equiv \frac{\|\vb*{L}^{(n)}_\text{2-lvl}-\vb*{L}^{(N)}_\text{2-lvl}\|_F}{\|\vb*{L}^{(N)}_\text{2-lvl}\|_F},
\end{align}
where $\|\cdot\|_F$ denotes the Frobenius norm, and $N$ is the total number of available snapshots. The norm $\|\cdot \|_D$ measures the normalized distance between the matrices constructed with $n$ and $N$ snapshots. The convergence of $\vb*{L}_\text{2-lvl}$ as more and more data is added is apparent from figure \ref{X_norm}.

\begin{figure}
		\centering
		\includegraphics[trim = 0mm 35mm 0mm 0mm, clip, width=1\textwidth]{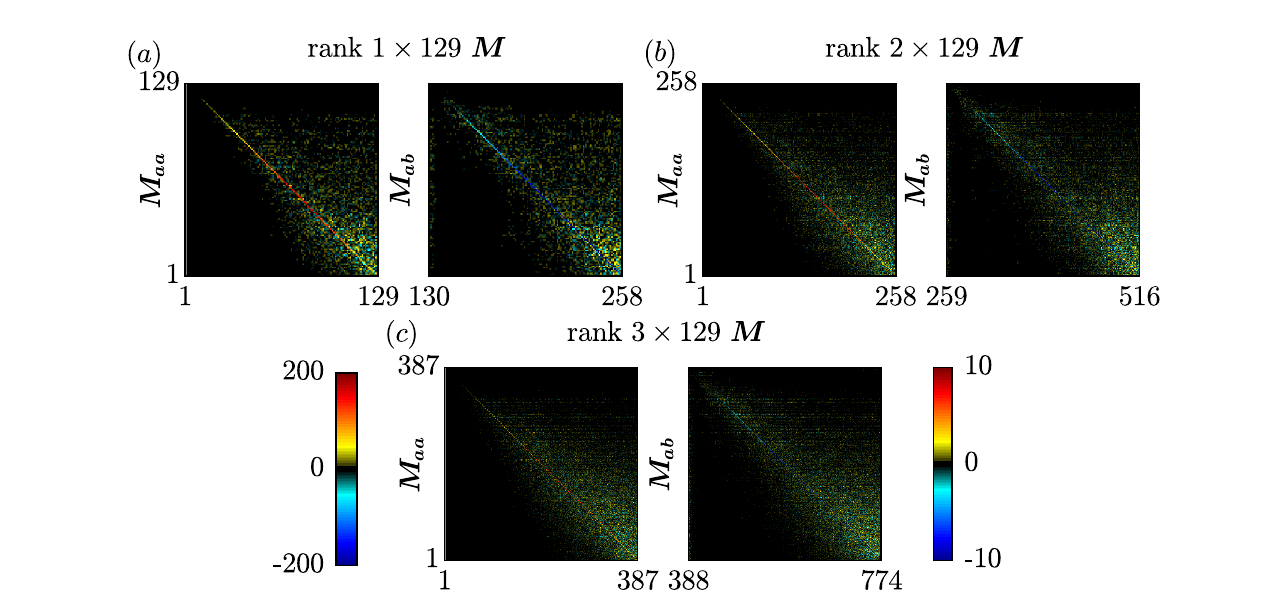}
        \caption{Matrices $\vb*{M}_{ab}$ (left column) and $\vb*{M}_{bb}$ (right column), computed from the least-squares problem (\ref{M0}) for models of different rank: ($a$) rank $1\times129$;  ($b$) rank $2\times129$; ($c$) rank $3\times129$.} \label{M_rank123}
	\end{figure}

Figure \ref{M_rank123} compares the left, $\vb*{M}_{ab}$ (left column), and the right, $\vb*{M}_{bb}$ (right column), square matrices that together form $\vb*{M}$, as defined in equation (\ref{M}),
for modal basis of three different sizes. 
The matrix $\vb*{M}_{ab}$ represents the cross--correlation information between the expansion and the forcing coefficients. 
Compared to the off-diagonal entries, a notable positive correlation is observed on the diagonal of the matrix $\vb*{M}_{ab}$. This suggests that the correlation between the expansion and the forcing coefficients at the same frequencies dominate.
The matrix $\vb*{M}_{bb}$ shows the cross-correlation between the forcing coefficients. This matrix, too, is diagonally-dominant; however, the autocorrelation of the forcing is negative. In the following discussion of figure \ref{a_f} below, we will show that this procedure of computing $\vb*{M}$ leads to an almost flat residue spectrum.

	\begin{figure} 
		\centering
        \includegraphics[trim = 0mm 0mm 0mm 0mm, clip, width=1\textwidth]{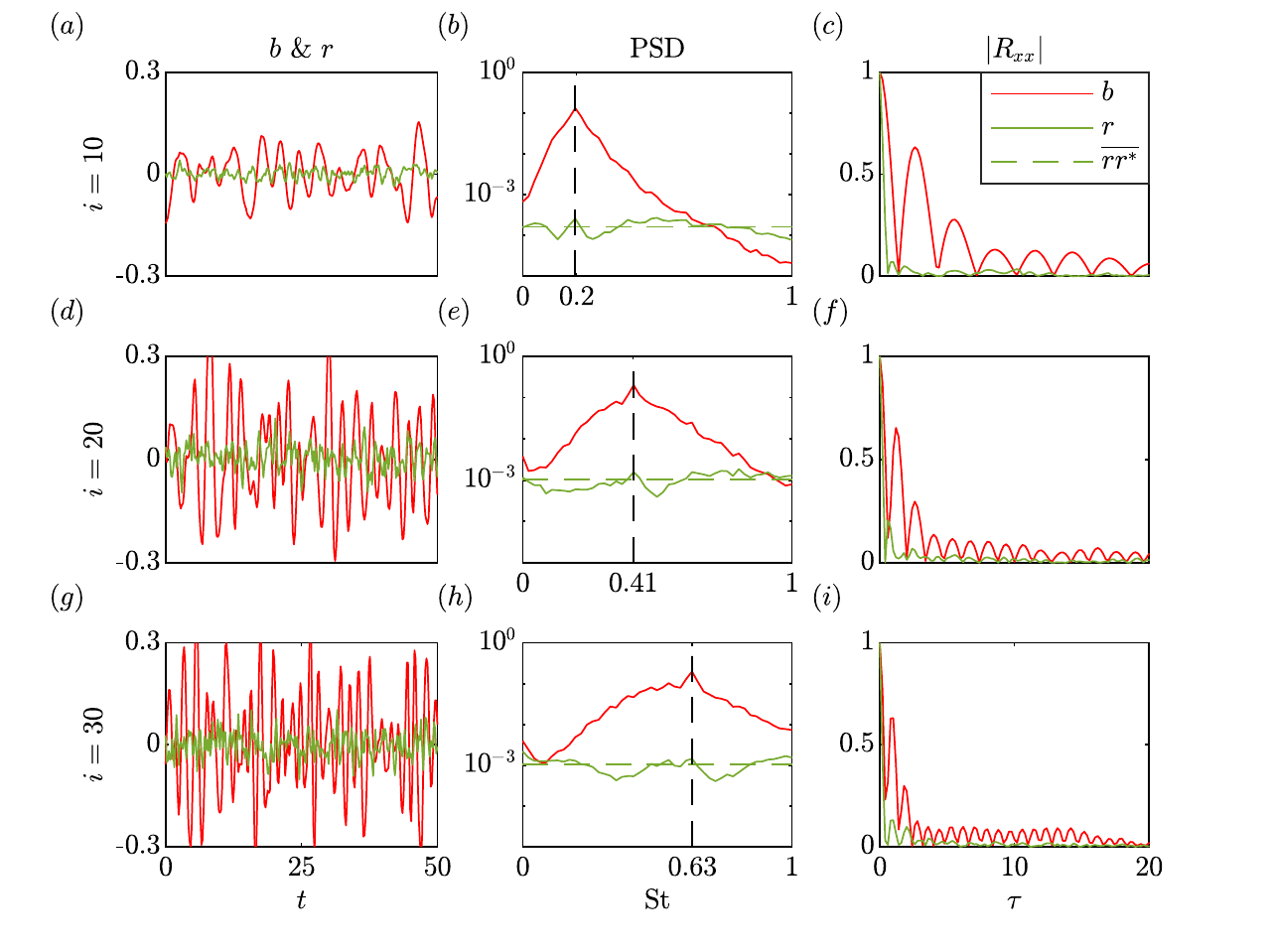}
        \caption{Time traces (left column), spectra (middle column) and autocorrelation estimates (right column) of the rank $1\times 129$ forcing coefficients, $\vb*{b}$ (red), and residue, $\vb*{r}$ (green), at three representative frequencies: $(a-c)$ $i=10$, or $St=0.2$; $(d-f)$ $i=20$, or $St=0.41$; $(g-i)$ $i=30$, or $St=0.63$. In (b,e,h), vertical dashed lines indicate the SPOD mode frequencies and horizontal dashed green lines the mean PSD of the residue.
        The absolute values of the normalized autocorrelation is shown in (c,f,i).}\label{a_f}
	\end{figure}

Following the steps outlined in \S \ref{residue}, we proceed with the modeling of the second-level residue. Recall that the proposed model closure hinges on the assumption that the highest-level residue can be modeled as random noise. This assumption is tested in figure \ref{a_f}, which examines the temporal evolution and the spectra of the forcing coefficients, $\vb*{b}(t)$, and the residue, $\vb*{r}(t)$. The rank 1$\times$129 case is shown as an example. It is observed that the residue is of significantly lower amplitude than the forcing.
Similar to the trend observed for the SPOD expansion coefficient in figure \ref{A_q}, the PSD of the forcing coefficients, shown in the right column, attains its maximum value at the corresponding mode frequency. For the frequencies at hand, a separation of at least two orders of magnitude between the maximum and minimum values of the PSD is found. The PSD of the residue, on the contrary, is significantly flatter.
To further quantify the white noise assumption,
	figure \ref{a_f}(c,f,i) show the corresponding (normalized) autocorrelations calculated as $R_{xx}(\tau) = \left<x(t)x(t+\tau)\right>$, where the ensemble average $\left<\cdot\right>$ is taken over the same blocks for the SPOD, see \S \ref{SPOD_jet}.
	The autocorrelations of the residue decay rapidly, whereas
	the forcing coefficients are correlated over 20 or more time units. This suggests that the residue can be modeled as white-in-time. Both the spectral flatness and the rapid drop of the autocorrelation motivate the truncation of the multi-level stochastic model at the second level.

% \begin{figure}
% 		\centering
% 		\includegraphics[trim = 0mm 0mm 0mm 0mm, clip, width=0.8\textwidth]{autocorrelation.pdf}
%         \caption{\Red{Autocorrelation functions $R_{xx}(\tau)$ for the rank $1\times 129$ forcing coefficients, $\vb*{b}$ (red), and residue, $\vb*{r}$ (green), at three representative frequencies: $(a)$ $i=10$, or $St=0.2$; $(b)$ $i=20$, or $St=0.41$ and $(c)$ $i=30$, or $St=0.63$. Autocorrelations are normalized by the corresponding magnitude at zero time-lag, respectively.}} \label{autocorrelation}
% 	\end{figure}

	\begin{figure} 
		\centering
        \includegraphics[trim = 5mm 10mm 15mm 5mm, clip, width=0.7 \textwidth]{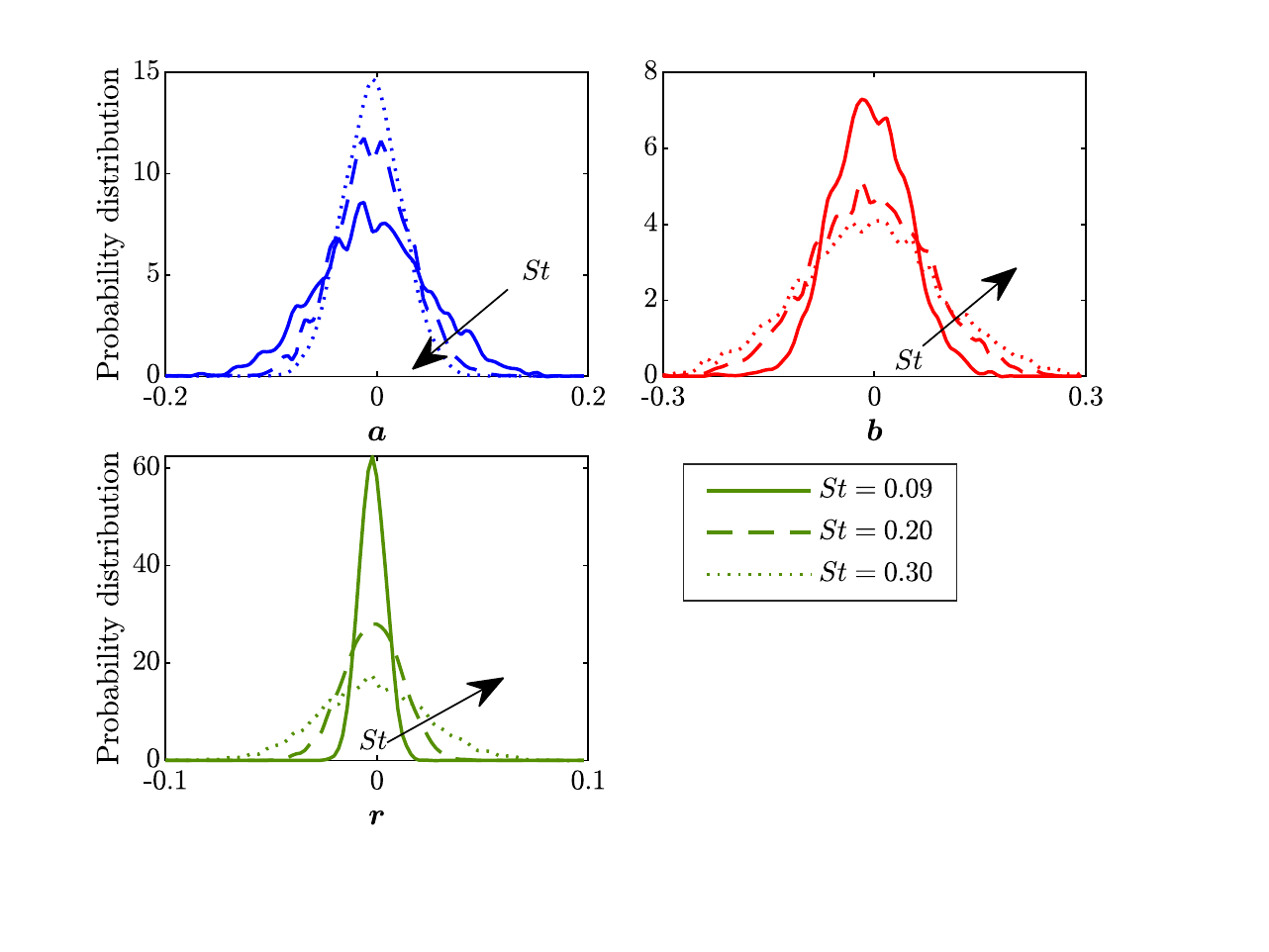}
        \caption{The probability distribution of rank 1$\times$129 state coefficients $\vb*{a}$ (blue), forcing coefficients $\vb*{b}$ (red) and the residue $\vb*{r}$ (green) at different frequencies: $St=0.09$ (solid), $St=0.20$, (dashed), $St=0.30$ (dotted). }\label{PDF_rank1}
	\end{figure}
To further motivate the proposed model truncation at the second level, we next examine the probability distributions of the rank-1$\times$129 coefficients $\vb*{a}$ and $\vb*{b}$, and the corresponding residue $\vb*{r}$ in figure \ref{PDF_rank1}. As desired, the probability distributions of residue at different frequencies are nearly Gaussian, which is clearly not the case for the mode and forcing coefficients. In accordance with the observations made in the context of figure \ref{a_f}, this suggests that $\vb*{r}$ may be modelled with components that are mutually correlated, but that are white-in-time. In the model, the correlation between the components of the residue is accounted for by the matrix $\vb*{G}$ (see equation (\ref{G}) that filters the white-in-time input ${\vb*{w}}$ to generate the process noise $\widetilde{\vb*{w}}$ that drives the final model.

\subsection{Subspace and flow field realizations} \label{result}
Consistent with the equation (\ref{sde_d}), we use the forward Euler method to march the SDE, equation (\ref{sde}), forward in time. The time step of the original data, $\Delta t=0.2$, is used to guarantee consistency with the modeling of the forcing and the residue.

	\begin{figure} 
		\centering
        \includegraphics[width=1\textwidth]{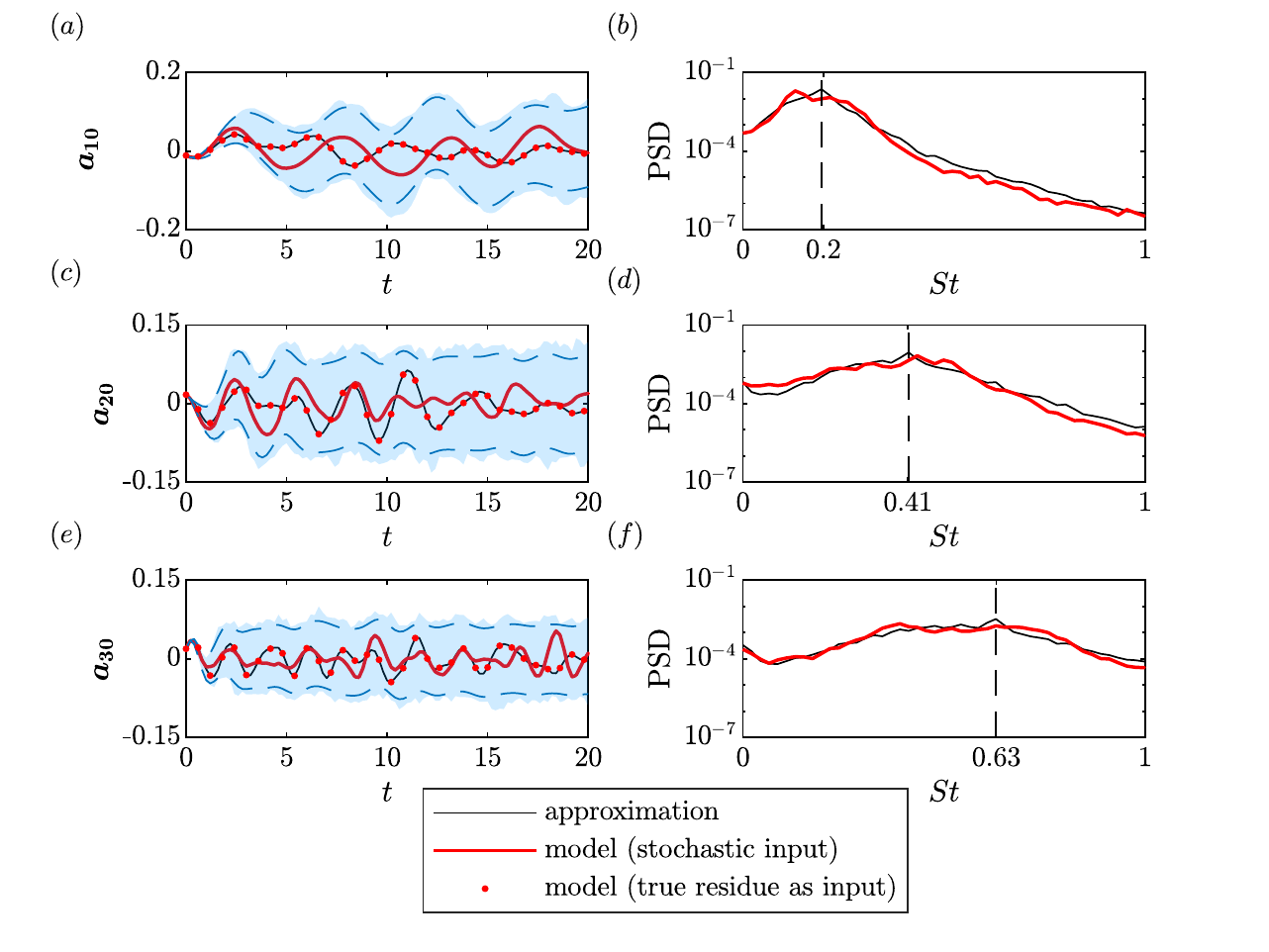}
        \caption{Time series and spectra for the rank 1$\times$129 approximation (black line, representing the original data) and a single realization of the rank 1$\times$129 model (red line) for three representative frequencies: (a,b) $St=0.2$; (c,d) $St=0.41$; (e,f) $St=0.63$. Red dots show the model output when the true residue taken from the data is used as input. It is indistinguishable from the data.
        Blue dashed lines mark the analytic $95\%$ confidence interval. The blue shaded area outlines a Monte Carlo envelope based on $10^4$ realizations of the stochastic model, for comparison.} \label{uncertainty} 
	\end{figure}
	
% After establishing that	
	
Figure \ref{uncertainty} shows the comparison between the rank 1$\times$129 approximation and a single realization of the rank 1$\times$129 model, as well as their power spectra at different frequencies. Starting from an initial condition taken from the data, this example of a realization of the stochastic model follows the initial transient dynamics of the data for approximately 3 time units. 
To demonstrate that the model accurately captures the known trajectory, we run the model without stochastic forcing and the true residue as the input $\vb*{r}$. It can be seen that the model output (red dots) and the LES trajectory (black line) are indistinguishable. 
A more rigorous approach to quantify the predictability of the model is use of the analytical 95\% confidence interval defined by equation (\ref{conf}), and Monte Carlo simulation. For the latter, $10^4$ realizations of the stochastic model, all starting from the same initial condition, were computed. It can be seen that the envelope of uncertainty closely follows the reference in the vicinity of $t=0$. 
As theoretically predicted in the context of equation (\ref{conf}), the region of uncertainty stays bounded for larger times, implying that the model is stable. A good agreement is also found between the power spectra of the coefficients of the model and the rank $1\times 129$ approximation. This observation confirms that the model preserves the direct correspondence between modes and frequencies inherent to SPOD.

		\begin{figure} 
		\centering
        \includegraphics[trim = 3mm 0mm 5mm 0mm, clip, width=1\textwidth]{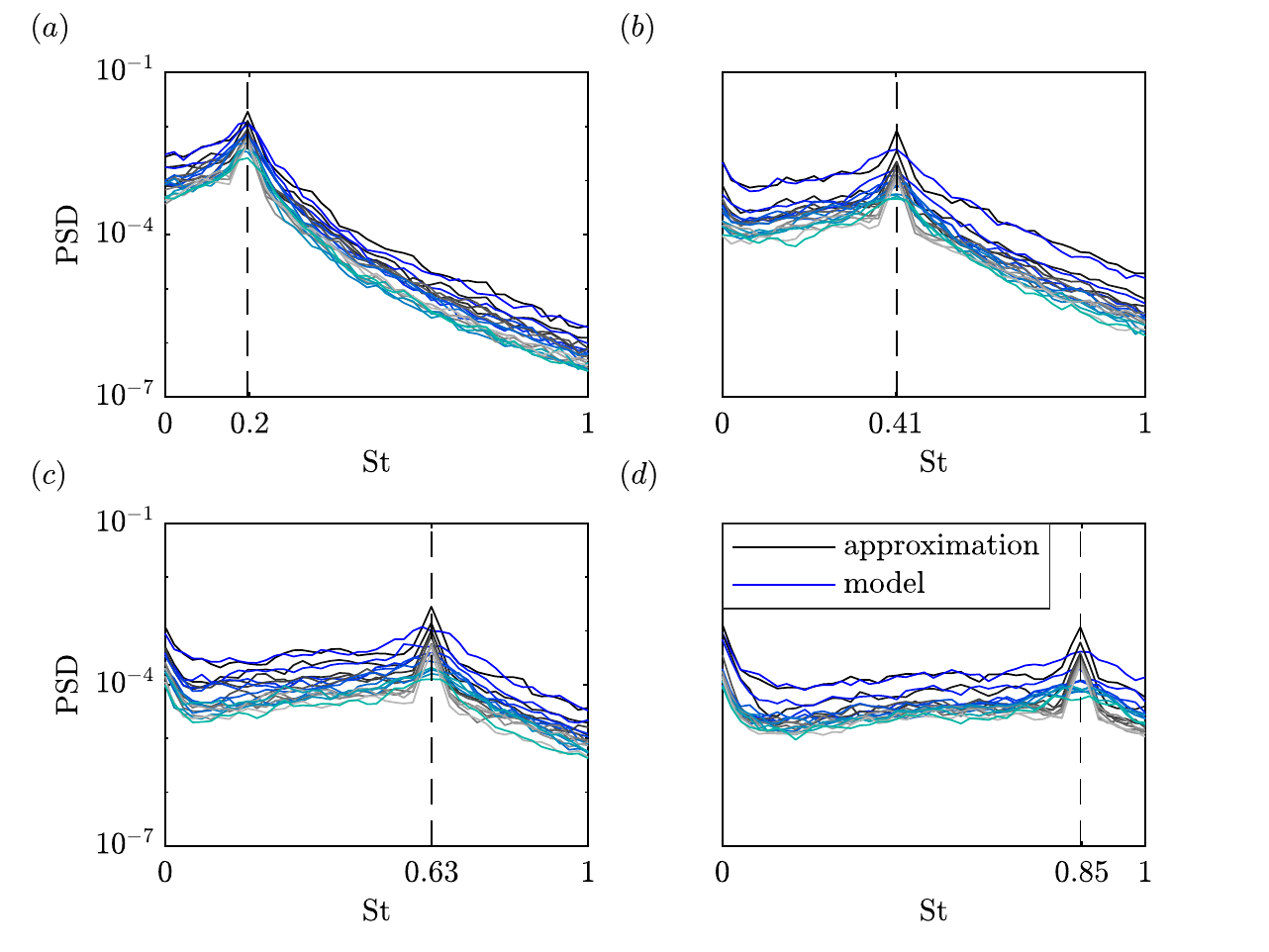}
        \caption{The power spectra of the first three state coefficients $\vb*{a}_i$ of rank 10$\times$129 approximation (black) and model (blue) at four representative frequencies: (a) $St=0.2$; (b) $St=0.41$; (c) $St=0.63$; (d) $St=0.85$ .} \label{psd_r10} 
	\end{figure}
	
Figure \ref{psd_r10} shows, analogously to figure \ref{uncertainty}(b,d,f), the power spectra of the approximation and model coefficients at different frequencies for the rank 10$\times$129 approximation. For all frequencies, the power spectra of the rank 10$\times$129 approximation and model follow the order of the SPOD eigenvalues. As for the rank 1$\times$129 model, a favourable agreement between approximation and model is observed. In summary, figures \ref{uncertainty} and \ref{psd_r10} demonstrate that both the dynamics and statistics of the state coefficients $\vb*{a}$ are well described by two-level models of different fidelity. More results for the rank 2$\times$129 and 3$\times$129 models are provided in appendix \ref{app}. 

	\begin{figure}
		\centering
        \includegraphics[trim = 0mm 0mm 0mm 5mm, clip, width=0.75\textwidth]{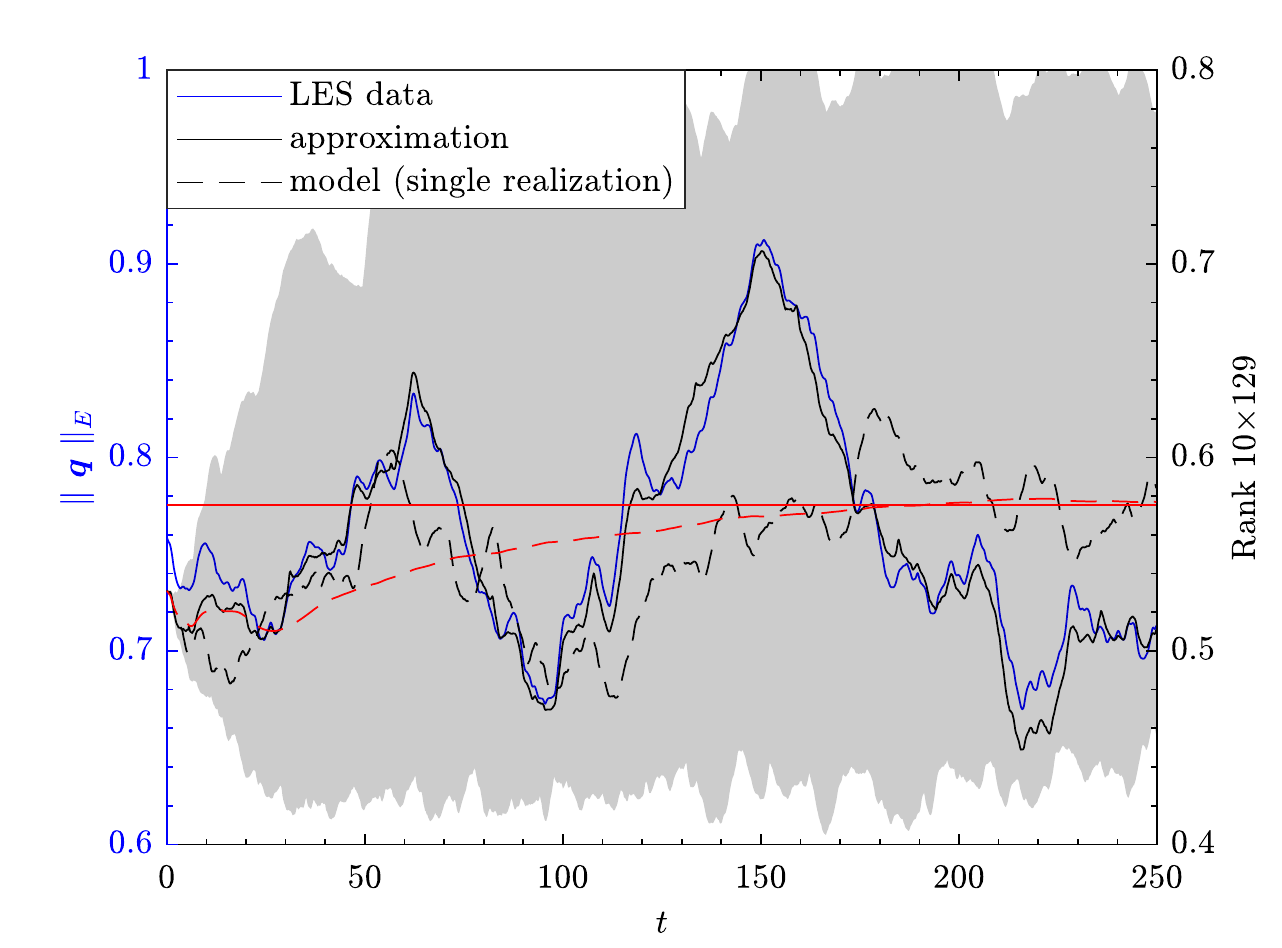}
        \caption{Comparison between the compressible energy norms of LES data (blue), rank 10$\times$129 approximation (black), and a single realization of the stochastic model (black dashed). The shaded area is a Monte Carlo envelope based on 2500 model evaluations.
        The ensemble average of Monte Carlo simulations (red dashed) and
        the long-time mean of rank 10$\times$129 approximation (red solid) are shown for comparison.   \label{comp_model}}
	\end{figure}
Having found that the spectral content of individual projection coefficients is represented well by the model, we now focus on the entire fluctuating flow field $\vb*{q}'$. Figure \ref{comp_model} compares the compressible energy norm of the LES data, the rank $10\times 129$ , and a single realization of rank $10\times 129$ model. Also indicated in gray is the region of uncertainty obtained from $2500$ Monte Carlo realizations of the model that start from the same initial condition. Two shifted axes are used to account for the constant offset between the oblique projection and model, on one hand, and the full data, on the other hand (see discussion of figure \ref{C_norm}). A favorable agreement of the general dynamics is observed between the original data a and the rank $10\times 129$ model. Starting from the initial condition of the data, the random realization shown here (first Monte Carlo sample) follows the general trend of the data for, arguably, up to 100 time units. The model uncertainty region show that significantly larger variations are of course possible, but that the model stays bounded within an expected range. As desired, the ensemble average of the Monte Carlo simulations converges to the long-time mean of reduced-rank approximation.

% To have a more straightforward view of the snapshot, we normalize the low-rank approximations and models to have the same root mean square of the LES data. For example, the normalized pressure is defined as 
% \begin{align}
%     \widetilde{p}=\frac{p}{\|\vb*{p}\|_\text{rms}}.
% \end{align}

	\begin{figure}
		\centering
        \includegraphics[trim = 5mm 42mm 5mm 8mm, clip, width=1\textwidth]{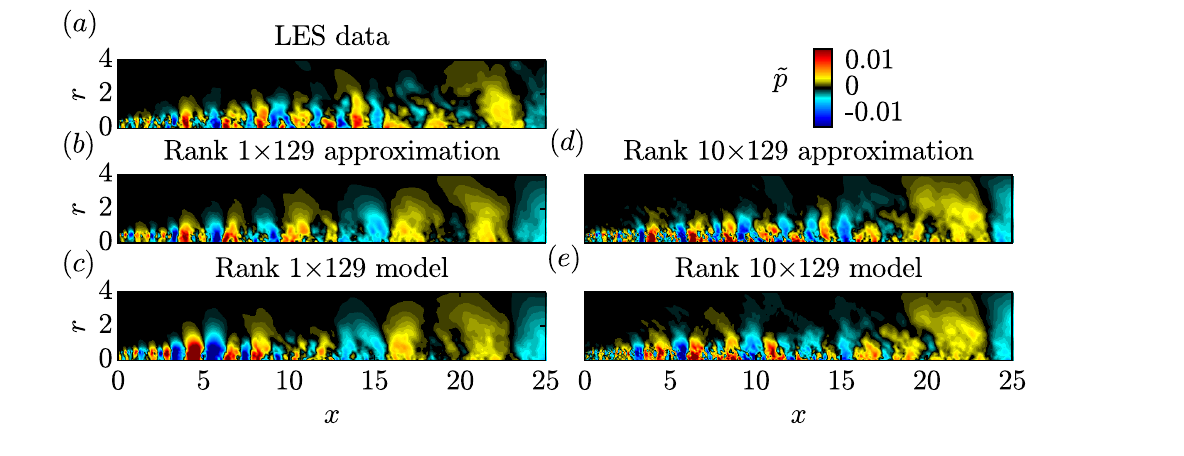}
        \caption{Comparison between pressure fields of the LES data, its low-rank approximations, and random realizations of the two-level model at $t=5$: (a) LES data; (b,c) rank 1$\times$129  and model; (d,e) rank 10$\times$129  and model. The pressure is normalized by its root mean square for comparability (see discussion of figure \ref{C_norm}).}\label{t=5}
	\end{figure}
After establishing that the approach yields a good model of the flow dynamics in terms of energy, we next compare the original, approximated and modeled flow fields. As examples, figure \ref{t=5} compares the instantaneous pressure fields at $t=5$ for rank 1$\times$129 and rank 10$\times$129. The same realization of the stochastic rank 10$\times$129 model as in figure \ref{comp_model} above (dashed line) is shown. From figure \ref{comp_model}, we expect that the model closely follows the approximation (and therefore the data) at this short time after its initialization. This can be clearly seen for both the rank 1$\times$129 and rank 10$\times$129 approximations and models in figure \ref{t=5}(b,c) and \ref{t=5}(d,e), respectively. As expected, the higher-rank approximation and model yield a more detailed picture of the flow. But even for rank 1$\times$129, many of the eminent features of the LES data are accurately captured. Due to the stochastic nature of the problem, this similarity fades for larger times. Instead, the stochastic model will yield a unique flow trajectory that can be interpreted as surrogate data that accurately reproduces the second-order statistics and dynamics of the input data. 

% 	\begin{figure}
% 		\centering
%         \includegraphics[width=0.8\textwidth]{r05_rank10.pdf}
%         \caption{Comparison between LES data, rank $10$  and model of pressure component at $r=0.5$ for different time. }\label{r05_rank10}
% 	\end{figure}

	\begin{figure}
		\centering
        \includegraphics[trim = 10mm 23mm 20mm 5mm, clip, width=1\textwidth]{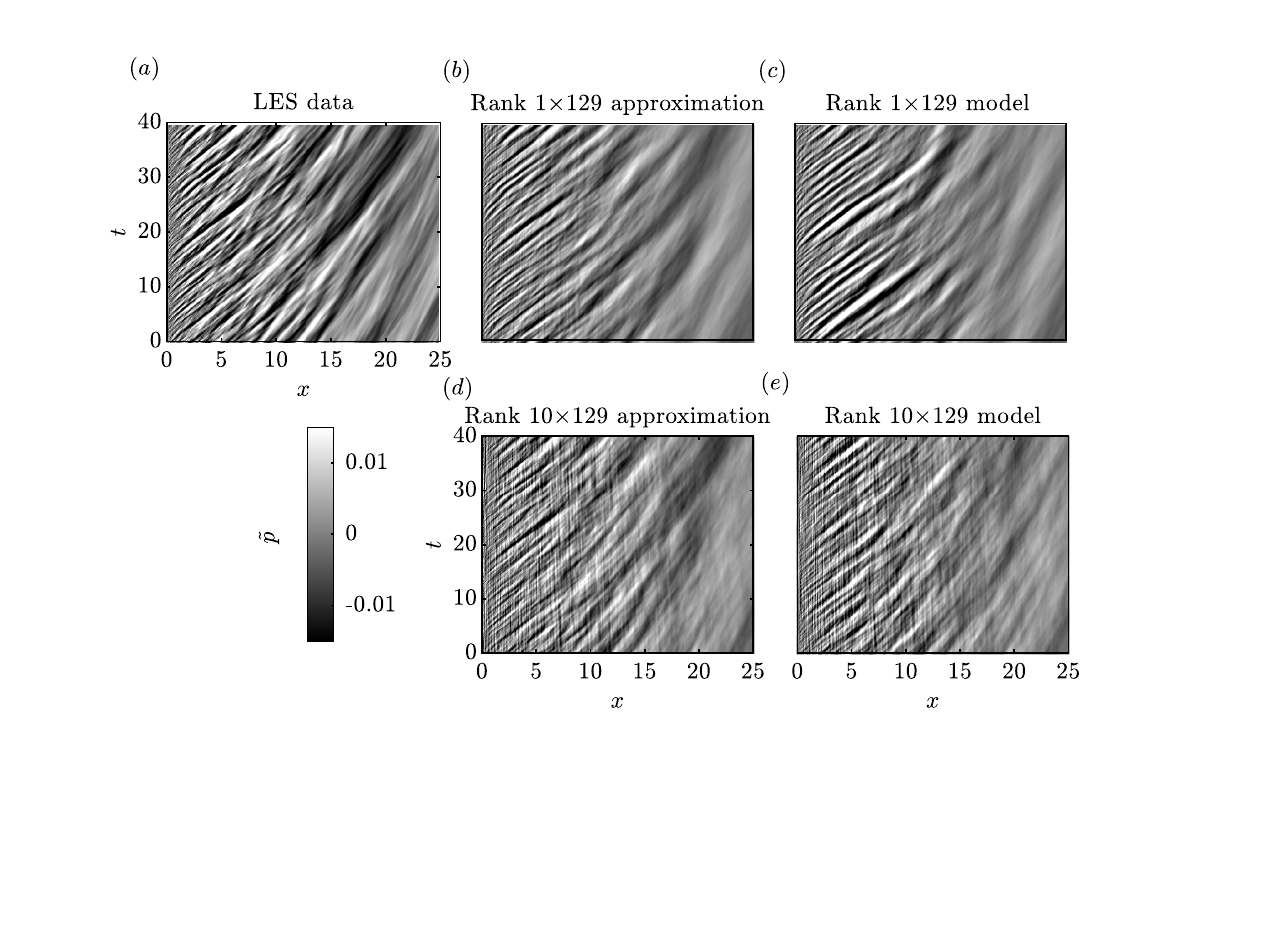}
        \caption{Comparison between LES data (a), rank $1\times129$  and model (b-c), rank $1\times129$  and model (d-e) of normalized pressure component at $r=0.5$ for different time.  }\label{r=0.5}
	\end{figure}

The temporal evolution of the fluctuating pressure field is investigated in terms of x-t diagrams at $r=0.5$ (along the lipline) in figure \ref{r=0.5}. The convective nature of the flow field becomes apparent from the diagonal pattern corresponding to the advection of the wavepackets previously seen in figure \ref{t=5}. The original LES data shown in figure \ref{r=0.5}(a) is compared to its low-rank approximations (middle) and the model output (right). It can be seen that the rank $1\times129$ model in figure \ref{r=0.5}(c) exhibits dynamics that are highly reminiscent of the rank $1\times129$ approximation and the full data. Similar observations are made for the rank $10\times129$ approximation and model in figure \ref{r=0.5}(d) and (e), respectively. We note that the vertical stripes in figure \ref{r=0.5}(d) stem from spatial aliasing, which occurs for subdominant modes at high frequencies. We emphasize that this issue is linked to the discretization of the data set, which was interpolated from the unstructured LES grid to a cylindrical grid for post-processing. The phenomenon is, however, accurately reproduced by the model, as can be seen in figure \ref{r=0.5}(e).

\newpage
	
		\begin{figure}
		\centering
        \includegraphics[trim = 0mm 10mm 0mm 20mm, clip,width=1\textwidth]{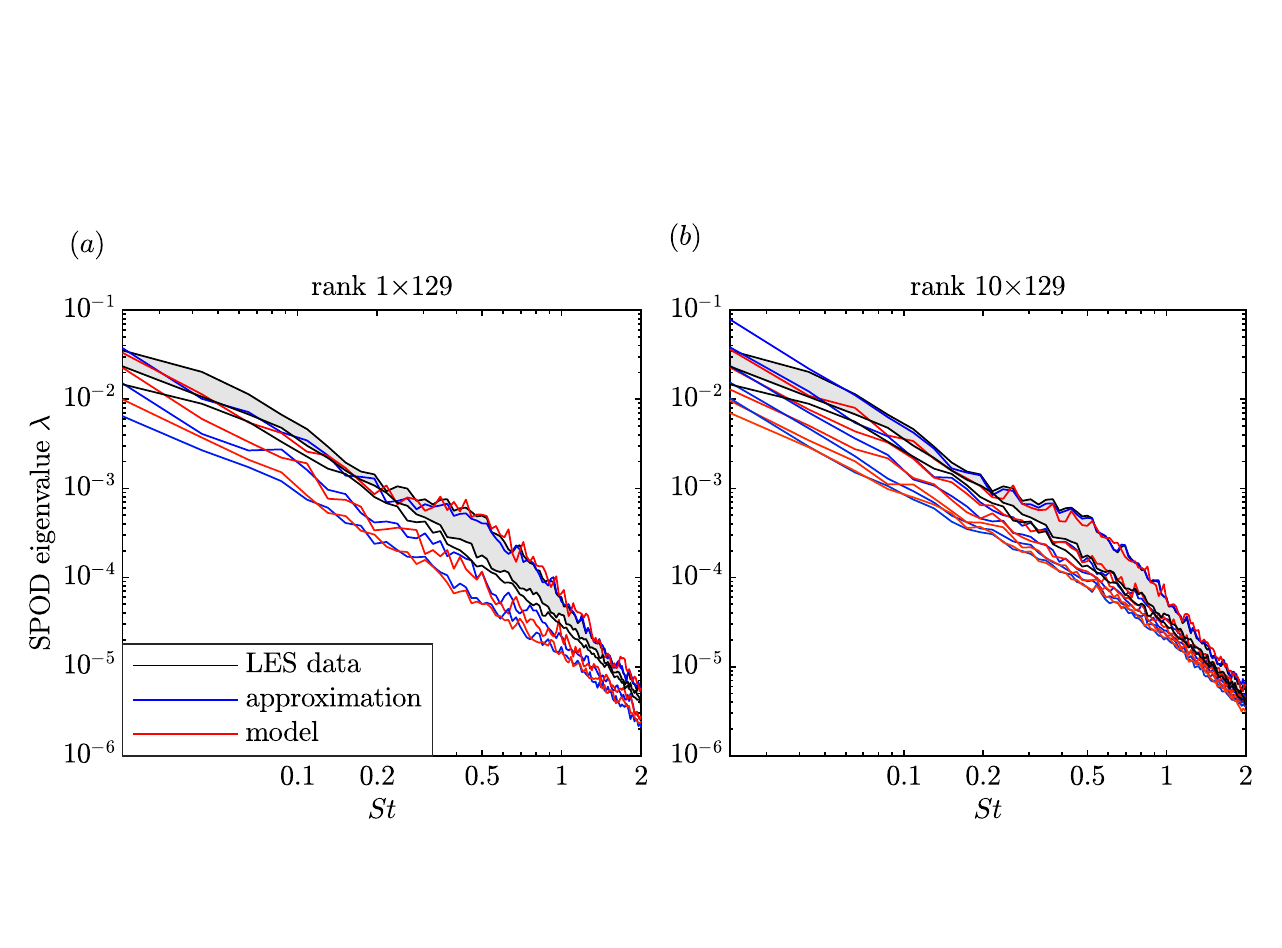}
\caption{Comparison of SPOD eigenvalue spectra of the LES data (black), low-rank approximations (blue) and 2--level models (red): (a) rank $1\times129$; (b) rank $10\times129$. The leading two eigenvalues are shown for the LES data (with shaded area between them), and the leading three eigenvalues for the approximations and models, respectively.}
\label{spectra_proj_model_r110}
	\end{figure}
	
After establishing that the model reproduces a surrogate flow field that is qualitatively very similar to the original data, we now focus on the flow statistics. An obvious choice is to use the SPOD for this purpose as well. In figure \ref{spectra_proj_model_r110}, we hence compare the SPOD eigenvalue spectra of LES data to both the rank $1\times 129$ and $10\times129$ approximations and models, respectively. For clarity, only the leading two eigenvalues are shown for the LES, and the leading $3$ eigenvalues for the rank $10\times129$ approximation and model. It can be seen that both models accurately reproduce the eigenvalue spectra of the LES data for all but the lowest frequencies. The rank $10\times129$ model accurately follows the $10\times129$ approximation at frequencies with $St>0.2$, but somewhat under-predicts the $10\times129$ approximation at lower frequencies. For the $1\times 129$ model, we observe an overall very good fit between approximation and model for the leading mode. In accordance with the conjectures drawn from figure \ref{r=0.5}, we conclude that the rank $1\times 129$ model produces an accurate surrogate flow field, both qualitatively and statistically.

\section{Conclusions}	\label{conclusion}

In this work, we propose stochastic two-level SPOD-Galerkin model for turbulent flows. The model is stable and produces surrogate data with accurate dynamics and statistics. The first level of the model is a forced Galerkin-ROM which propagates the SPOD-expansion coefficients in time. The SPOD-expansion coefficients are obtained by oblique projection of the data onto the SPOD basis. Consistent with this procedure, the system dynamics matrix of the ROM is obtained by projecting the discretized governing equations onto a low-dimensional SPOD basis. Here, we consider the linearized compressible Navier-Stokes equations. The offset between the linear approximation and the true state evolution is interpreted as a forcing of this linear system. The second level of the model is a stochastically forced system for this forcing. This closure is inspired by the purely data-driven linear inverse model by \citet{penland1996stochastic,penland1989random} and its generalization, the linear multi-level regression model by \citet{kravtsov2005multilevel,kondrashov2005hierarchy}. The basic idea behind these models is to achieve closure by inflating the state with additional levels until the residue can be modeled as white noise. In the purely data-driven model of \citet{kravtsov2005multilevel}, this criterion was met at the third level. For the turbulent jet governed by the compressible Navier-Stokes equations, we show that closure is achieved at the second level if a physical model is used describe the linear dynamics at the first level.

The proposed model is inspired by recent progress in flow modeling based on the resolvent operator, and the correspondence between resolvent and SPOD modes \citep{towne2018spectral} under the assumption of white noise forcing. A possible extension of this work is hence to substitute the SPOD basis by a resolvent basis, which only requires knowledge of mean flow. Another potential venue to improve the model is to further offset differences between the reduced-order representation and the true data, for example by using the ROM Error Surrogates (ROMES) method by 
\citet{drohmann2015romes}, which implements Gaussian-process regression to construct error surrogates.

\vspace{5mm}
\noindent{\bf Acknowledgements.} 

\vspace{5mm}
\noindent{\bf Declaration of Interests.} The authors report no conflict of interest.

%  \section*{Acknowledgments}

%%%%%%%%%%%%%%%%%%%%%%%%%%%%%%%%

\appendix

\newpage

\section{Stochastic SPOD-Galerkin two-level model for rank 2$\times$129 and rank 3$\times$129 cases}\label{app}

		\begin{figure} 
		\centering
        \includegraphics[trim = 5mm 15mm 10mm 0mm, clip, width=1\textwidth]{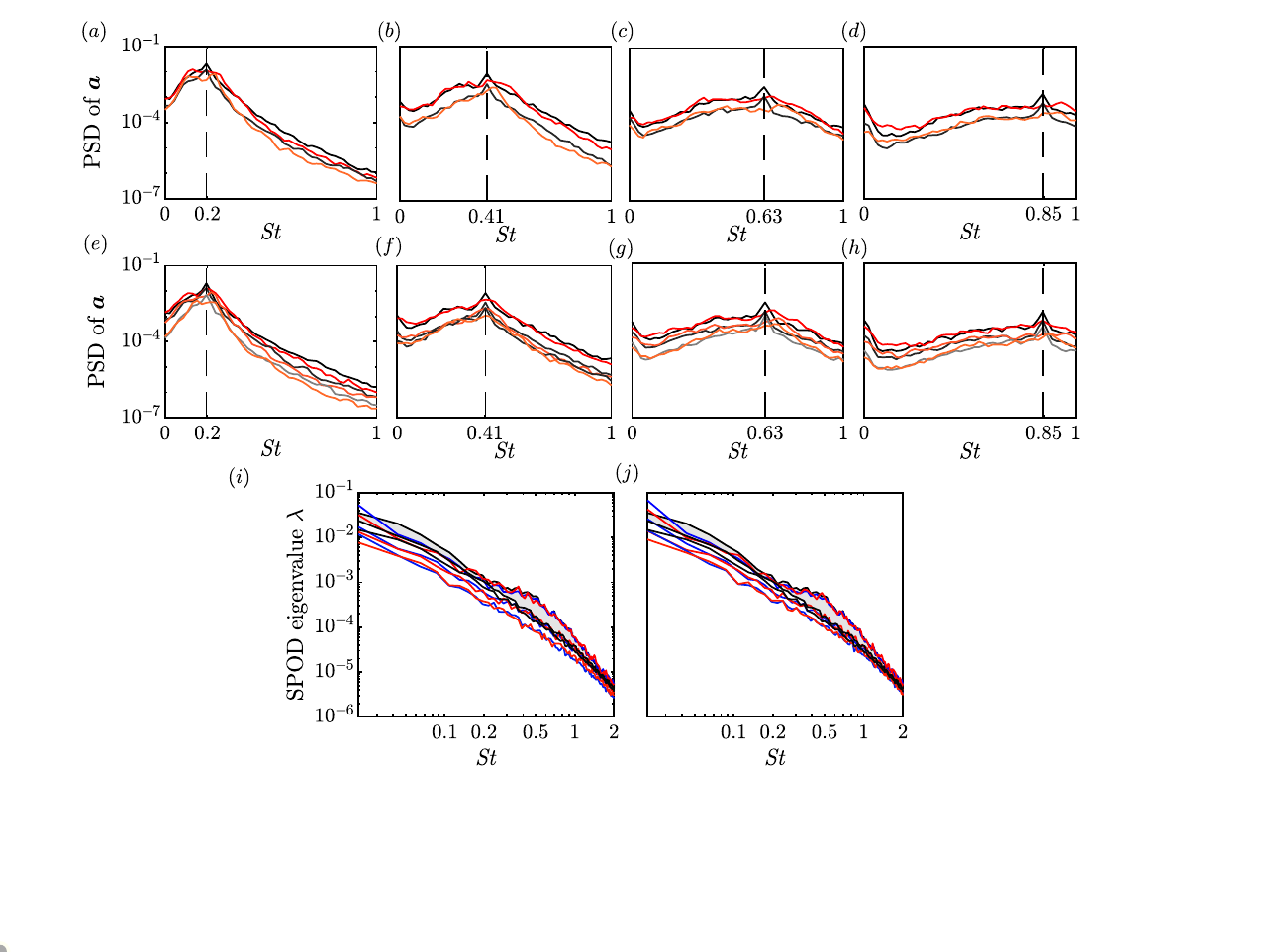}
        \caption{
        Power spectra of the state coefficients $\vb*{a}_i$ 
        of low-rank approximations (red) and 2--level models (black) of rank 2$\times$129 (first row) and rank 3$\times$129 (second row) at four representative frequencies: (a,e) $St=0.2$; (b,f) $St=0.41$; (c,g) $St=0.63$; (d,h) $St=0.85$. Comparison of SPOD eigenvalue spectra of the LES data (black), low-rank approximations (blue) and 2--level models (red): (i) rank $2\times129$; (j) rank $3\times129$. Compare figures \ref{uncertainty}, \ref{comp_model} and \ref{spectra_proj_model_r110}.} \label{psd_r2} 
	\end{figure}

	    This appendix reports the additional results for the rank 2$\times$129 and rank 3$\times$129 cases, which were omitted in section \ref{result} for brevity. Figure \ref{psd_r2}(a-d) show the comparison between the power spectra of the state coefficients $\vb*{a}_i$ and the rank 2$\times$129 expansion coefficients at different frequencies. Figure \ref{psd_r2}(e-h) report the corresponding results for the rank 3$\times$129 case. Good agreements between the approximations and models are observed in both cases. The corresponding SPOD eigenvalue spectra shown in figure \ref{psd_r2}(i-j) show that 
	    both models accurately reproduce the eigenvalue spectra of the LES data for a wide range of frequencies. From these observations and the favourable results obtained for the 1$\times$129 baseline model, as previously reported in figures \ref{uncertainty} and \ref{spectra_proj_model_r110}, it concluded that subdominant SPOD modes are not required in the modal expansion.

\bibliographystyle{jfm}
\bibliography{references}

\end{document}